\documentclass[sigplan,preprint,nonacm]{acmart}
\usepackage{booktabs}
\usepackage{float}
\usepackage{multirow}
\usepackage{pifont}
\usepackage{xurl}
\usepackage{graphicx}
\usepackage{xcolor}
\definecolor{hl}{RGB}{232,238,246}
\definecolor{acc}{RGB}{31,78,121}
\usepackage{algorithm}
\usepackage{algorithmic}
\usepackage{pgfplots}
\usetikzlibrary{patterns,decorations.pathreplacing}
\pgfplotsset{compat=1.17}
\usepackage{microtype}
\usepackage{tcolorbox}
\newcommand{\gbs}[1]{#1\,GB/s}
\newtcolorbox{findingbox}{
  colback=black!4, colframe=black!40, boxrule=0.5pt, arc=1.5pt,
  left=5pt, right=5pt, top=3pt, bottom=3pt, boxsep=0pt,
  before skip=4pt plus 2pt, after skip=4pt plus 2pt}
\newcommand{\finding}[2]{%
\begin{findingbox}\small\textbf{Finding (#1):} #2\end{findingbox}}
\newcommand{\MachRead}{517}          

\newcommand{\MlxDefault}{53}
\newcommand{\PtDefaultLo}{61}
\newcommand{\PtDefaultHi}{82}
\newcommand{\BestCopy}{134}
\newcommand{\AdoptPt}{516}           
\newcommand{\ResOurs}{523}           
\newcommand{\ResMlxDense}{497}       

\newcommand{\CubeMapDevStream}{489}
\newcommand{\CubeMapDevOwn}{526}
\newcommand{\CubeMapRtOwn}{265}
\newcommand{\CubeMapRtStream}{169}
\newcommand{\CubeCopyDevStream}{76}
\newcommand{\CubeCopyRtStream}{60}
\newcommand{\CubeCopyDevOwn}{57}
\newcommand{\CubeCopyRtOwn}{54}
\newcommand{\CubeRefResident}{523}

\newcommand{\CubeRefStock}{61}

\newcommand{\CubeMapDevStreamIQR}{[484, 491]}
\newcommand{\CubeMapDevOwnIQR}{[519, 528]}
\newcommand{\CubeMapRtOwnIQR}{[265, 269]}
\newcommand{\CubeMapRtStreamIQR}{[166, 169]}
\newcommand{\CubeCopyDevStreamIQR}{[74, 79]}
\newcommand{\CubeCopyRtStreamIQR}{[58, 60]}
\newcommand{\CubeCopyDevOwnIQR}{[56, 58]}
\newcommand{\CubeCopyRtOwnIQR}{[54, 55]}
\newcommand{\CubeRefResidentIQR}{[520, 526]}

\newcommand{\CubeRefStockIQR}{[55, 61]}

\newcommand{\CubeCopyLo}{54}           
\newcommand{\CubeCopyHi}{76}

\newcommand{\MlxAdopted}{517.5}                     
\newcommand{\MlxResident}{506.1}                     
\newcommand{\MlxAdoptPct}{+0.7}                        
\newcommand{\MlxAdoptCI}{[-1.6\%, +9.3\%]}          

\newcommand{\ReplayMapped}{0.121}     
\newcommand{\ReplayCopy}{1.219}       
\newcommand{\ReplayOwnRt}{1.262}      
\newcommand{\ReplayStock}{4.109}      

\newcommand{\SizeSmallMap}{28.9}      
\newcommand{\SizeBigMap}{550}         
\newcommand{\SizeCopyCap}{87}         
\newcommand{\SizeCrossMB}{23}         
\newcommand{\SizeSmallRatio}{4.4}     
\newcommand{\WsBigMap}{578}           

\newcommand{\AudCorrection}{0.1}       
\newcommand{\AudSteadyPass}{2}
\newcommand{\AudResident}{477}         
\newcommand{\AudMmap}{25.2}
\newcommand{\AudMalloc}{413}
\newcommand{\AudMallocVram}{2.15}      
\newcommand{\AudProcs}{7}

\newcommand{\LcppStockTps}{2.82}     
\newcommand{\LcppAdoptTps}{3.42}     
\newcommand{\LcppSpeedup}{1.21}      
\newcommand{\LcppStockWs}{8.13}      
\newcommand{\LcppAdoptWs}{4.20}      
\newcommand{\LcppStockRatio}{1.86}
\newcommand{\LcppAdoptRatio}{0.96}
\newcommand{\LcppModelGB}{4.36}
\newcommand{\LcppShards}{3}
\newcommand{\ApuHeapCeil}{12.5}      

\newcommand{\LcppDiscStock}{146.9}
\newcommand{\LcppDiscAdopt}{3.79}
\newcommand{\LcppDiscSlowdown}{39}   

\newcommand{\CapPressFree}{10.3}     
\newcommand{\CapPressModel}{8.4}     
\newcommand{\CapPressStockWs}{14.7}
\newcommand{\CapPressAdoptWs}{8.1}
\newcommand{\CapPressStockWall}{58.8}
\newcommand{\CapPressAdoptWall}{22.4}

\newcommand{\PolAppleMap}{109.9}
\newcommand{\PolAppleRes}{109.5}
\newcommand{\PolAppleCopy}{900}      
\newcommand{\PolApuMap}{135.3}      
\newcommand{\PolApuRes}{154.0}      
\newcommand{\PolGhMap}{340}
\newcommand{\PolGhCopy}{335}
\newcommand{\PolDiscMap}{97.6}
\newcommand{\PolDiscRes}{5.7}
\newcommand{\PolApuCopy}{679}        
\newcommand{\PolDiscCopy}{82.2}      
\newcommand{\PolTieBand}{5}          

\newcommand{\PairSessMapped}{518.9}    
\newcommand{\PairSessResident}{520.3}
\newcommand{\PairSessPure}{522.6}
\newcommand{\PairDiffMedian}{+0.16}    
\newcommand{\PairMapped}{516.2}                 
\newcommand{\PairResident}{517.8}                 
\newcommand{\PairPure}{522.2}                 
\newcommand{\PairDiffCI}{[-0.66\%, +0.48\%]}    
\newcommand{\PairPureCI}{[-1.41\%, -0.23\%]}    

\newcommand{\OrdStreamCell}{0.26}     
\newcommand{\OrdEventCell}{0.54}      
\newcommand{\OrdDrainCell}{2.27}      

\newcommand{\OrdEventCI}{[-0.14, +0.95]}   
\newcommand{\OrdDrainCI}{[+1.41, +2.63]}   

\newcommand{\AblFull}{0.115}          
\newcommand{\AblNoCthree}{2.7}        
\newcommand{\AblNoCtwo}{4.5}          
\newcommand{\AblNoCone}{17.0}         
\newcommand{\AblStock}{32.7}          

\newcommand{\FactFull}{0.04}          
\newcommand{\FactCthreeOff}{3.46}
\newcommand{\FactCtwoOff}{5.62}

\newcommand{\FactConeOff}{12.38}

\newcommand{\FactAllOff}{17.90}
\newcommand{\FactFullIQR}{[-0.23, 0.23]}
\newcommand{\FactCthreeOffIQR}{[3.28, 3.73]}
\newcommand{\FactCtwoOffIQR}{[5.18, 5.68]}

\newcommand{\FactConeOffIQR}{[12.15, 12.65]}

\newcommand{\FactAllOffIQR}{[17.53, 18.22]}
\newcommand{\FactNoCone}{12.3}        
\newcommand{\FactNoCtwo}{5.6}         
\newcommand{\FactNoCthree}{3.4}       

\newcommand{\OrdInStream}{0.33}                  
\newcommand{\OrdEvent}{0.44}                  
\newcommand{\OrdShared}{3.46}                  
\newcommand{\OrdDrain}{5.80}                  
\newcommand{\OrdProcesses}{8}                 
\newcommand{\OrdInStreamCI}{[0.24, 0.47]}     
\newcommand{\OrdEventLocalCI}{[0.40, 0.56]}   
\newcommand{\OrdSharedCI}{[3.14, 3.67]}       
\newcommand{\OrdDrainMechCI}{[5.66, 6.02]}    
\newcommand{\OrdEvMinusStream}{+0.17}         
\newcommand{\OrdEvMinusStreamCI}{[+0.02, +0.21]}  

\newcommand{\KthreeSpineStock}{2.62}
\newcommand{\KthreeSpineCopy}{1.34}
\newcommand{\KthreeSpineForeign}{5.96}
\newcommand{\KthreeSpineMapped}{0.35}
\newcommand{\KthreeTokenBefore}{4.14}
\newcommand{\KthreeTokenAfter}{1.87}
\newcommand{\KthreeMatrices}{1253}   
\newcommand{\KthreeFiles}{2506}      
\newcommand{\KthreeSpineGiB}{52.8}   
\newcommand{\KthreeSpineGB}{56.7}    
\newcommand{\KthreeWireGBs}{137}     
\newcommand{\KthreeFirstTouch}{1.37}
\newcommand{\KthreeMapOnly}{0.96}       
\newcommand{\KthreeWireOnly}{0.41}      

\newcommand{\QwenBytes}{71.48}      
\newcommand{\QwenPublic}{7.14}       
\newcommand{\QwenPublicLo}{5.76}
\newcommand{\QwenPublicHi}{7.20}
\newcommand{\QwenExtLo}{4.30}
\newcommand{\QwenExtHi}{6.81}
\newcommand{\QwenStockRatio}{7.6}    
\newcommand{\QwenExt}{6.13}          
\newcommand{\QwenStock}{0.94}        
\newcommand{\QwenPublicRate}{510}    
\newcommand{\QwenExtRate}{438}       
\newcommand{\QwenStockRate}{67}      

\newcommand{\CapRounds}{8}           
\newcommand{\CapStockParse}{32.59}   
\newcommand{\CapStockCopy}{4.51}     
\newcommand{\CapStockTTFT}{37.95}    
\newcommand{\CapRebindLoad}{0.21}    
\newcommand{\CapRebindFirst}{6.75}   
\newcommand{\CapRebindTTFT}{6.97}    
\newcommand{\CapCopyTTFT}{8.42}      
\newcommand{\CapTok}{6.9}            
\newcommand{\CapLoadRatio}{6.4}      
\newcommand{\CapLoadRatioCI}{[5.4, 11.2]}  
\newcommand{\CapAdoptOnlyStartup}{1.6}     
\newcommand{\CapColdRounds}{3}
\newcommand{\CapColdStockTTFT}{32.1}
\newcommand{\CapColdRebindTTFT}{6.9}
\newcommand{\CapColdCopyTTFT}{7.2}
\newcommand{\CapColdLoadRatio}{5.6}        

\newcommand{\GhResident}{3063}
\newcommand{\GhLink}{376}
\newcommand{\GhOverlap}{355}
\newcommand{\GhMmap}{350}
\newcommand{\GhHostAlloc}{346}
\newcommand{\GhMallocLo}{15}
\newcommand{\GhMallocHi}{20}

\newcommand{\GhMigLo}{0.25}
\newcommand{\GhMigHi}{0.92}
\newcommand{\GhSeventyTwo}{2.48}     

\newcommand{\DiscResident}{815}
\newcommand{\DiscMapped}{48}

\newcommand{\ApuMapped}{41.3}

\newcommand{\AirChip}{M3}
\newcommand{\AirRam}{8}
\newcommand{\AirModel}{Qwen2.5-1.5B}
\newcommand{\AirW}{1.29}             
\newcommand{\AirMatrices}{140}
\newcommand{\AirMap}{56.0}           
\newcommand{\AirResident}{56.9}
\newcommand{\AirCopy}{8.4}
\newcommand{\AirResFits}{4/5}        
\newcommand{\AirPublicTps}{25.4}     

\newcommand{\AirCopyTps}{0.90}
\newcommand{\AirAdoptRatio}{28}      
\newcommand{\AirDmMap}{59.3}         
\newcommand{\AirDmResident}{60.7}
\newcommand{\AirDmCopy}{7.7}
\newcommand{\AirDmResFits}{5/5}      
\newcommand{\AirDmPublicTps}{30.6}   
\newcommand{\AirDmCopyTps}{0.99}
\newcommand{\AirDmAdoptRatio}{31}    

\newcommand{\ShareBigFile}{65.5}           
\newcommand{\ShareBigAnon}{61.2}           
\newcommand{\ShareBigFileEvict}{30.9}      
\newcommand{\ShareBigMapTokHi}{5.5}        
\newcommand{\ShareBigMapTokLo}{3.6}        
\newcommand{\ShareBigResTok}{0.08}         
\newcommand{\AirBigW}{2.74}          
\newcommand{\AirEvictAfter}{0}       

\newcommand{\ApuW}{4.68}             
\newcommand{\ApuMapDep}{34.6}        
\newcommand{\ApuResDep}{30.4}        
\newcommand{\ApuCopyDep}{6.9}        
\newcommand{\ApuRatioDep}{1.14}      
\newcommand{\ApuResCached}{53.5}     
\newcommand{\ApuResFullSet}{30.2}     
\newcommand{\ApuStageInflation}{1.8} 
\newcommand{\GranBytes}{4096}       
\newcommand{\ApuGranOne}{1.174}      
\newcommand{\ApuGranThree}{1.174}    

\newcommand{\PrivMapGB}{60}          
\newcommand{\PrivAnonGB}{58}         
\newcommand{\PrivWireSlowdown}{10}   
\newcommand{\PrivTputLoss}{3}        

\newcommand{\ReferenceMachine}{Apple M5~Max}          
\newcommand{\ReferenceRamGB}{128}                     
\newcommand{\ReferenceOS}{macOS~26.6}                 
\newcommand{\TorchVersion}{2.13}                      
\newcommand{\MlxSurveyVersion}{0.29.3}                
\newcommand{\MlxAdoptVersion}{0.32.0}                 
\newcommand{\ProtocolRows}{12288}                     
\newcommand{\ProtocolCols}{7168}                      
\newcommand{\ProtocolMatrixMB}{88.1}                  
\newcommand{\ProtocolMatrices}{12}                    
\newcommand{\WarmupPasses}{12}                        
\newcommand{\CILevel}{95}                            
\newcommand{\SurveyProcesses}{15}                     
\newcommand{\PairProcesses}{10}                       
\newcommand{\FactorialProcesses}{8}                   
\newcommand{\FactEffCone}{12.49}

\newcommand{\FactEffCtwo}{3.72}

\newcommand{\FactEffCthree}{1.86}

\newcommand{\FactEffConeCtwoCI}{[-0.15, 0.17]}

\newcommand{\FactEffConeCthreeCI}{[-0.33, 0.18]}
\newcommand{\FactEffCtwoCthree}{1.79}
\newcommand{\FactEffCtwoCthreeCI}{[1.67, 2.20]}

\newcommand{\FactEffThreeWayCI}{[-0.31, 0.24]}
\newcommand{\GhTieGapPct}{1.4}                        
\newcommand{\PolThetaLo}{2}                          
\newcommand{\PolThetaHi}{20}
\newcommand{\DiscBus}{57}                             
\newcommand{\ProducerLines}{226}                      
\newcommand{\ProdRateStd}{526.1}                     
\newcommand{\ProdRateExt}{521.8}                     
\newcommand{\ProdRateDiffPct}{0.83}                 
\newcommand{\KernelMaxT}{8}                           
\newcommand{\OneCopy}{1}
\newcommand{\TwoCopies}{2}

\newcommand{\KthreeName}{Kimi~K3}
\newcommand{\KthreeParamsT}{2.8}                      
\newcommand{\KthreeLayers}{93}                        
\newcommand{\KthreeRoutedLayers}{92}                  
\newcommand{\KthreeExperts}{896}                      
\newcommand{\KthreeTopK}{16}                          
\newcommand{\KthreeExpertPoolTB}{1.446}               
\newcommand{\KthreeExpertTokenGiB}{24.0}              
\newcommand{\KthreeBytePct}{86}                       
\newcommand{\KthreeArithmeticPct}{14}                 
\newcommand{\KthreeLiveRuns}{3}                       
\newcommand{\KthreeStageSpeedup}{7.5}                 
\newcommand{\KthreeAdoptOnly}{3.8}                    
\newcommand{\KthreeTokenSpeedup}{2.2}                 
\newcommand{\KthreeBadSlowdown}{2.3}                  
\newcommand{\KthreeTaxSaved}{2.27}                   
\newcommand{\KthreeTokenRest}{1.52}                  

\newcommand{\QwenModel}{Qwen2.5-72B}
\newcommand{\QwenProcesses}{8}                        
\newcommand{\QwenResident}{7.23}                      
\newcommand{\QwenResidentRate}{517}                   
\newcommand{\QwenResProcs}{4}                         
\newcommand{\QwenStockCI}{[0.85, 0.97]}               
\newcommand{\QwenFpSixteenGB}{145}                    
\newcommand{\CompatibilityModel}{Qwen2.5-7B}
\newcommand{\CompatMappedRSS}{0.5}                    
\newcommand{\CompatStockRSS}{17.9}                    
\newcommand{\CompatMappedTps}{23.0}                   
\newcommand{\CompatStockTps}{22.9}                    
\newcommand{\CompatPairs}{5}                          
\newcommand{\CapacityModel}{Qwen2.5-32B}
\newcommand{\CapacityCheckpointGB}{65}                

\newcommand{\ApuPlatform}{AMD Ryzen~7 9700X APU}      
\newcommand{\ApuRuns}{4}                              
\newcommand{\GhPlatform}{NVIDIA GH200}
\newcommand{\GhHbmGB}{96}                             
\newcommand{\GhCpuGB}{480}                            
\newcommand{\GhBigGB}{119}                            
\newcommand{\DiscretePlatform}{NVIDIA RTX~5070~Ti}
\newcommand{\HotSetLo}{75}                            
\newcommand{\HotSetHi}{80}                            

\newcommand{\WorkingSetPoints}{4}                     
\newcommand{\WorkingSetMinGB}{2}
\newcommand{\WorkingSetMaxGB}{16}
\newcommand{\WorkingSetCoords}{%
  (2.0,548.2) (4.1,558.6) (8.0,570.8) (16.0,578.0)}
\newcommand{\SizeMappedCoords}{
  (1.048576,28.9) (4.194304,77.8) (16.777216,328.1)
  (33.554432,441.5) (88.080384,523.4) (134.217728,550.1)}
\newcommand{\SizeCopyCoords}{
  (1.048576,6.5) (4.194304,24.5) (16.777216,73.5)
  (33.554432,82.3) (88.080384,86.7) (134.217728,83.0)}

\newcommand{\PolAppleChoice}{map (tie)}
\newcommand{\PolAppleObserved}{adopted}
\newcommand{\PolApuChoice}{map}
\newcommand{\PolApuObserved}{adopted}
\newcommand{\PolGhChoice}{map (tie)}
\newcommand{\PolGhObserved}{tied}
\newcommand{\PolDiscChoice}{resident}
\newcommand{\PolDiscObserved}{rejected}

\newcommand{\TaxTOne}{5.31}        
\newcommand{\TaxTTwo}{3.66}        
\newcommand{\TaxTFour}{1.88}        
\newcommand{\TaxTEight}{1.50}        
\newcommand{\TaxTSixteen}{1.25}        
\newcommand{\TaxTMax}{16}
\newcommand{\TaxTokLo}{425}        
\newcommand{\TaxTokHi}{609}        

\begin{document}
\title{The Ingestion Tax: Adopting File-Backed Weights in Tensor Frameworks}
\author{Yuan Si}
\affiliation{%
  \institution{University of Waterloo}
  \city{Waterloo}
  \country{Canada}
}

\author{Yufeng Lin}
\affiliation{%
  \institution{Independent Researcher}
  \city{Macau}
  \country{China}
}

\author{Daming Li}
\affiliation{%
  \institution{Independent Researcher}
  \city{Mountain View}
  \country{USA}
}

\author{Jialu Zhang}
\authornote{Corresponding author.}
\affiliation{%
  \institution{University of Waterloo}
  \city{Waterloo}
  \country{Canada}
}
\begin{abstract}
Open-weight models can occupy a middle capacity regime: active weights fit in
DRAM as cached file pages, but a second framework-owned copy does not fit
or must be refilled as layers run, so low-batch decode rereads the
weights every token.  On integrated and coherent-memory systems those file
pages are already GPU-readable, yet ordinary loading paths copy them into
framework allocations before use.  We call this copy the
\emph{ingestion tax}.

We present \emph{file-backed weight adoption}: a framework-independent
producer maps each tensor with \textsc{map\_shared}, wraps the pages as a
no-copy GPU buffer, and exports a DLPack capsule that PyTorch or MLX imports
as ordinary storage.  Zero-copy import alone is insufficient: the
implementation must also keep activations accelerator-resident and
establish ordering on the GPU; an adopter that omits both runs a
dense decode stage $\KthreeBadSlowdown\times$ slower than stock in the
live system.  With both in place, adoption
removes the tax: the public route reaches
\AdoptPt\,GB/s versus \MlxDefault--\PtDefaultHi\ for the default
constructors, matches the identical kernel over
resident storage (\PairDiffCI, paired), and is within 1.3\% of a
resident control on a matched
\QwenModel\ (\QwenPublic\ vs.\ \QwenResident\,tok/s).

At the same throughput, the weights remain clean, shared, evictable file
pages: $N$ processes decode from one mapped copy where resident loading
creates $N$ copies (at capacity, \ShareBigMapTokHi\ vs.\
\ShareBigResTok\,tok/s), and a \CapacityCheckpointGB\,GB checkpoint
cuts time to first token by \CapLoadRatio$\times$ versus stock loading.  In
\KthreeName, a 2.8T-parameter MoE, the dense int8 spine stage
falls from \KthreeSpineStock\ to \KthreeSpineMapped\,s per token
(\KthreeStageSpeedup$\times$; \KthreeAdoptOnly$\times$ from storage
alone).  The same mechanism improves llama.cpp by
\LcppSpeedup$\times$ at half the footprint on an AMD APU, falls inside
the \PolTieBand\% selection band of overlapped streaming on a
capacity-exceeding GH200 workload, and is \LcppDiscSlowdown$\times$
slower across PCIe.  The deployment rule follows memory topology: adopt
file pages only where the GPU can already read them.
\end{abstract}

\maketitle
\section{Introduction}
A model whose active weights exceed accelerator-owned capacity can still fit
in host DRAM and the page cache.  Low-batch decode then rereads each active
matrix per token.  Integrated GPUs and coherent CPU--GPU links allow those rereads to access
host-resident pages directly: the checkpoint's file pages already occupy a
memory domain the GPU can read, so a kernel need not move them to distinct
physical memory.

Tensor-framework loaders nevertheless make that move in software:
constructor or transfer paths allocate framework-owned storage and copy
file-backed bytes into it, and a streaming loader refills it
as layers run.  We call the transfer the \emph{ingestion tax}.  Unlike
offload from storage or a copy into discrete-GPU memory, it creates a
second software-owned representation in the same GPU-readable domain.
This consumes bandwidth and capacity without changing physical placement.

Standalone runtimes such as llama.cpp can expose mapped model storage to
Metal while controlling command submission~\citep{llamacpp}.  A tensor
framework must also preserve activation placement, ordering, aliasing,
lifetime, and mutation semantics.  DLPack and MLX's conditional zero-copy
Metal import~\citep{mlxdlpackdoc} alone do not make a checkpoint mapping a
fast, correctly ordered framework tensor.  An adopter that removes
ingestion this way but round-trips activations through the host and
dispatches outside the framework runs a dense decode stage
$\KthreeBadSlowdown\times$ slower than stock in the live system: the
synchronization it
adds costs more than the copy it removes (\S\ref{sec:motiv}).

This paper develops \emph{file-backed weight adoption}: map a tensor with
\textsc{map\_shared}, wrap the pages as a no-copy GPU buffer, export a
DLPack capsule, and import it as ordinary framework storage.  The design spans
operating-system and architecture boundaries.  On the operating-system side,
mapping semantics, page-cache residency, and clean-page reclaim decide what
storage is represented: adoption makes VM-managed file pages ordinary
accelerator storage that remains shared and evictable under existing kernel
policy.  On
the architecture side, memory topology (unified, coherent-link, or
discrete) decides whether reading that storage in place is profitable at
all.  Accordingly, we provide an execution contract and causal
decomposition for framework integration, a topology-driven placement rule,
and a public mechanism that connects the two layers.

\begin{enumerate}
\item \textbf{A cross-layer execution contract and causal decomposition.}
We separate three requirements: adopt the mapped pages (C1), retain
activations in the accelerator-visible domain (C2), and establish
dependencies on the GPU (C3).  A randomized factorial experiment removes
each condition while using the same kernel, tensor shapes, allocator,
and host path.  Removing C1, C2, or C3 adds $+\FactNoCone$, $+\FactNoCtwo$,
and $+\FactNoCthree$\,ms per pass, and the full $2^3$ model resolves one
interaction: ordering overhead is measurable only when activations remain
accelerator-resident.  A campaign across \OrdProcesses\ processes
separates the ordering mechanisms themselves: cost begins where a signal
becomes host-visible, not where the queue changes owner (\S\ref{sec:conditions}).

\item \textbf{A public adoption path with end-to-end impact.}  A
framework-independent producer turns aligned file mappings into versioned
Metal DLPack tensors.  The public PyTorch path reaches \AdoptPt\,GB/s under
the matched protocol, and MLX reaches \MlxAdopted\,GB/s through a capsule
from the same producer.  In \KthreeName, a packed kernel over adopted
storage takes the dense stage from \KthreeSpineStock\ to
\KthreeSpineMapped\,s, a $\KthreeStageSpeedup\times$ overall
speedup---the storage change alone contributes
\KthreeAdoptOnly$\times$ relative to the same packed kernel over
copied int8 (\KthreeSpineCopy\,s)---and cuts per-token latency from
\KthreeTokenBefore\ to \KthreeTokenAfter\,s.  A matched \QwenModel\ evaluation reaches
\QwenPublic\,tok/s through adoption versus \QwenStock\,tok/s through
per-use ingestion, within 1.3\% of a load-once resident control
(\QwenResident\,tok/s) while the weights stay reclaimable.  Rebinding
the parameters of stock \texttt{transformers} modules leaves the
model code unchanged and preserves probe logits bitwise, and a
capacity-scale
\CapacityModel\ cuts time to first token by \CapLoadRatio$\times$
versus the stock loader; the storage change alone contributes
\CapAdoptOnlyStartup$\times$, paired per round on identical files.

\item \textbf{A topology- and capacity-aware deployment rule.}  The same
loader decision improves llama.cpp by \LcppSpeedup$\times$ and removes its
duplicate model footprint on an \ApuPlatform, falls within the
\PolTieBand\% policy band of overlapped streaming for a capacity-exceeding
\GhPlatform\ workload, and is rejected on an \DiscretePlatform\ because
every mapped read crosses PCIe.  Full-working-set endpoint probes
select the observed outcome on all four platforms, and on a held-out
Apple machine one session exercises the capacity gate and the other
the tie band; in both, mapping is the faster of the two endpoints run
end to end.
\end{enumerate}

\section{Regime and execution contract}\label{sec:regime}

\subsection{The motivating deployment}\label{sec:motiv}
Large open-weight models can run locally by streaming from storage the
weights that do not fit in memory.  \KthreeName, a
\KthreeParamsT-trillion-parameter mixture of experts (\KthreeLayers\
layers, \KthreeExperts\ routed experts, \KthreeTopK\ selected in each of
\KthreeRoutedLayers\ routed layers), runs on one \ReferenceMachine\ with
\ReferenceRamGB\,GB of unified memory using this strategy: each token reads
\KthreeExpertTokenGiB\,GiB of experts out of a \KthreeExpertPoolTB\,TB
pool (the routed subset cached on local SSD), plus a dense int8
``spine'' of \KthreeMatrices\ matrices totaling \KthreeSpineGiB\,GiB
that fits the page cache.  Although the workload streams experts from SSD,
the measurements identify a larger cost elsewhere.  An identity ablation
that preserves byte reads while removing layer arithmetic attributes \KthreeBytePct\% of decode time to the byte path and \KthreeArithmeticPct\% to arithmetic.
The spine accounts for the largest share of byte-path time even though its
bytes already sit in DRAM: on every token the framework re-ingests them
into its own allocation, widening the packed int8 checkpoint to fp32 as
it copies, and that stage alone takes \KthreeSpineStock\,s of the
\KthreeTokenBefore\,s token.

A mapped implementation must do more than remove that copy: it must
remain within the framework's execution path.  Mapping the weights while dispatching from a private queue outside
the framework and round-tripping activations through the host removes the
copy but runs the stage $\KthreeBadSlowdown\times$ \emph{slower}
(\KthreeSpineForeign\ versus \KthreeSpineStock\,s; 3 runs,
Table~\ref{tab:k3}): the synchronization it adds costs more
than the copy (\S\ref{sec:conditions}).  A packed-int8 kernel over
weights adopted into the framework's own execution path instead runs the
stage in \KthreeSpineMapped\,s and the token in \KthreeTokenAfter\,s
(\S\ref{sec:eval}).

The resulting \KthreeTaxSaved\,s reduction per
token is larger than the \KthreeTokenRest\,s of the token that remain
outside the spine stage for attention and sampling; the
\KthreeExpertTokenGiB\,GiB expert stream (25.8\,GB, or 1.56\,s at
the local SSD's measured 16.5\,GB/s) overlaps the token's other work
rather than occupying its own interval.  Building a second representation of bytes the GPU can
already read is therefore the largest measured per-token cost in
this configuration.  The routed experts remain disk-bound and are
unchanged by this work; a companion study evaluates the expert-residency
tier itself~\citep{expertcache}.

\begin{figure*}[t]
\centering
\includegraphics[width=0.8\textwidth]{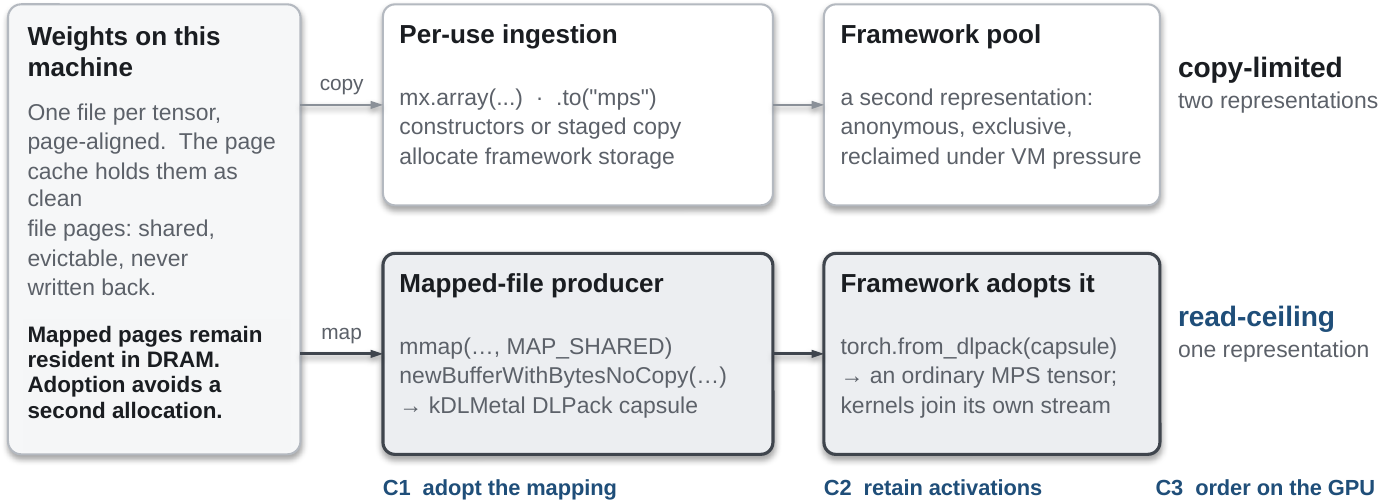}
\caption{Two byte paths from the same clean file pages: per-use ingestion builds a second, framework-owned representation; adoption republishes the pages as ordinary framework storage.  The bottom row names the three conditions C1--C3 of Table~\ref{tab:contract} that the adoption path must satisfy.}
\label{fig:overview}
\end{figure*}

\subsection{Where the tax applies, and what it is not}
The regime is bounded on both sides.  If the active weights fit
framework-owned storage, one copy amortizes over many tokens; if the
active set does not fit the page cache, storage traffic dominates and no
in-memory byte path helps.  Between those endpoints, the active files
are GPU-readable but a second allocation either does not fit or must be
refilled as layers run.  Installed memory alone is an insufficient capacity criterion because the copy
competes with the mapped checkpoint, activations, the OS, and dtype
materialization.  Adoption avoids the duplicate allocation and leaves the
tensor's storage as clean file pages that the OS may share or evict.

The stock path can combine three costs that must not be conflated.
\emph{Ingestion} copies file-backed bytes into framework-owned storage.
\emph{Representation expansion} converts a packed checkpoint into a wider
kernel format.  \emph{Execution handoff} moves activations or waits on the
host when a streamed kernel runs outside the framework's dependency graph.
File-backed adoption removes ingestion, but it does not by itself provide a
packed kernel or correct ordering.

\begin{table}[tb]
\centering\small\setlength{\tabcolsep}{2.8pt}
\begin{tabular}{@{}lp{0.66\columnwidth}@{}}
\toprule
Condition & Requirement \\
\midrule
\textbf{C1: adopt} & The kernel reads the mapped file pages; no per-use
framework copy is created. \\
\textbf{C2: retain} & Inputs and outputs remain accelerator-visible across
layer calls; activations do not round-trip through host arrays. \\
\textbf{C3: order on GPU} & Dependencies are established in the framework
stream or with an on-GPU event; the host does not drain outstanding work. \\
\bottomrule
\end{tabular}
\caption{The execution contract.}
\label{tab:contract}
\end{table}

Figure~\ref{fig:overview} contrasts the two byte paths, and
Table~\ref{tab:contract} states the required execution contract as
three conditions.  A standalone runtime normally owns all three.  In a tensor
framework, the mapped pages begin outside the allocator and scheduler.  The
core design problem is therefore to make those pages ordinary framework
storage.

\section{Characterizing the ingestion tax}\label{sec:tax}

\subsection{Matched protocol}
Every Apple-silicon rate in this section uses one task: a single-token
int8 GEMV over \ProtocolRows$\times$\ProtocolCols\ weights
(\ProtocolMatrixMB\,MB) with fp16 row scales.  A pass visits
\ProtocolMatrices\ page-cache-resident files and brackets the batch with
one synchronization; the protocol discards \WarmupPasses\ warm-up passes
and reports the median.  Each primary cell is replicated across
\SurveyProcesses\ processes with arm order randomized, and dispersion
is the IQR across
process medians.  The system is an \ReferenceMachine\
(\ReferenceRamGB\,GB unified memory, \ReferenceOS,
PyTorch~\TorchVersion, MLX~\MlxSurveyVersion~\citep{mlx}; MLX
\MlxAdoptVersion\ for
the \texttt{mx.from\_dlpack} measurements below).  Rates count weight bytes once, the traffic common to
the compared kernels.
Table~\ref{tab:tax} places the routes against the protocol's
\MachRead\,GB/s pure-read reference (a ceiling of this 12-dispatch
protocol, not a machine limit; \S\ref{sec:scale}): constructors reach
\MlxDefault--\PtDefaultHi\,GB/s, the strongest staged copy \BestCopy,
and adoption \AdoptPt.

\begin{table}[tb]
\centering\small\setlength{\tabcolsep}{3.4pt}
\begin{tabular}{@{}lrr@{}}
\toprule
How weights reach the kernel & GB/s & live copies \\
\midrule
MLX constructor & \MlxDefault & \TwoCopies \\
PyTorch \texttt{.to("mps")} & \PtDefaultLo--\PtDefaultHi & \TwoCopies \\
pipelined, double-buffered copy & \BestCopy & \TwoCopies \\
\midrule
\textbf{public DLPack adoption} & \textbf{\AdoptPt} & \textbf{\OneCopy} \\
\bottomrule
\end{tabular}
\caption{How weights reach the kernel: one int8 GEMV over
\ProtocolMatrices\ page-cache-resident \ProtocolMatrixMB\,MB matrices on
the \ReferenceMachine, against the protocol's \MachRead\,GB/s
pure-read reference.
\emph{Live copies} counts the weight representations resident during
decode: the mapped file pages plus any framework-owned copy.}
\label{tab:tax}
\end{table}

With weights already in each framework's pool, MLX reaches
\ResMlxDense\,GB/s and our packed-int8 kernel reaches
\CubeRefResident, both near the pure-read reference.  The constructor gap
therefore opens before GEMV execution.  Residency measurements show the
same distinction (\S\ref{sec:eval}, Table~\ref{tab:sharing}):
constructor paths grow anonymous framework-owned storage,
while adoption leaves the clean mapped pages as the only weight
representation.

\subsection{What DLPack already provides}
MLX documents \texttt{mx.array} as copying; it documents
\texttt{mx.asarray} and \texttt{mx.from\_dlpack} as capable of zero-copy
Metal import when the underlying buffer is reusable and non-private
\citep{mlxdlpackdoc}.  PyTorch exposes memory-sharing DLPack import and
custom MPS shaders \citep{torchdlpackdoc,torchmpsshader}.  These interfaces
do not directly construct a Metal capsule from a file mapping; our producer
supplies that step.  Capsules from the same producer reach
\AdoptPt\,GB/s through \texttt{torch.from\_dlpack} and
\MlxAdopted\,GB/s through \texttt{mx.from\_dlpack}, versus
\MlxResident\,GB/s for the same MLX~\MlxAdoptVersion\ kernel over its
resident allocation;
the median per-pass paired mapped-minus-resident difference is
\MlxAdoptPct\% with interval \MlxAdoptCI, while the arm medians differ
by $+2.25$\%.  Mutation-visibility checks establish storage
identity; file checksums and kernel output oracles confirm that the
adopted tensor reads the mapped bytes.

\subsection{Mapped storage reads at resident rate}\label{sec:pairedcamp}
A rotated paired campaign measures adoption against the identical kernel
over resident weights.  Across \PairProcesses\ independent processes, the
medians are \PairMapped, \PairResident, and \PairPure\,GB/s for the
mapped, resident, and pure-read arms (this campaign's pure-read arm
sits 1\% above the survey's \MachRead\,GB/s reference).  The median per-process paired
mapped-minus-resident difference is \PairDiffMedian\% with interval
\PairDiffCI\ (the ratio of the raw arm medians is $-0.31$\%), and the
mapped-minus-pure-read interval is
\PairPureCI\ (all intervals in this paper are \CILevel\% bootstrap
intervals over process medians).  The first interval bounds any
mapped-versus-resident loss to approximately one percent; the second
shows a small GEMV overhead relative to the pure-read
arm.

\subsection{Scale, granularity, and batch}\label{sec:scale}
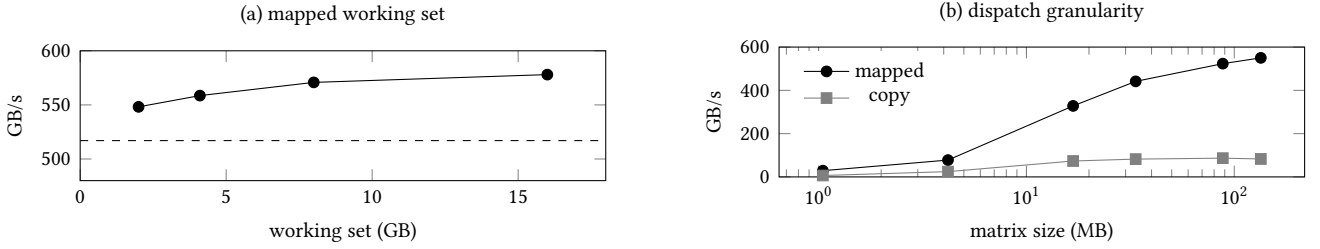
\begin{figure*}[t]
\centering
\begin{minipage}[t]{0.48\textwidth}
\centering
\begin{tikzpicture}
\begin{axis}[width=\linewidth,height=3.3cm,xlabel={working set (GB)},
  ylabel={GB/s},xmin=0,xmax=18,ymin=480,ymax=600,
  tick label style={font=\footnotesize},label style={font=\footnotesize},
  title style={font=\footnotesize},title={(a) mapped working set}]
\addplot[mark=*] coordinates {\WorkingSetCoords};
\addplot[dashed,domain=0:18] {\MachRead};
\end{axis}
\end{tikzpicture}
\end{minipage}\hfill
\begin{minipage}[t]{0.48\textwidth}
\centering
\begin{tikzpicture}
\begin{axis}[width=\linewidth,height=3.3cm,xlabel={matrix size (MB)},
  ylabel={GB/s},xmode=log,ymin=0,ymax=600,ytick={0,200,400,600},
  tick label style={font=\footnotesize},label style={font=\footnotesize},
  title style={font=\footnotesize},title={(b) dispatch granularity},
  legend style={font=\footnotesize,at={(0.03,0.97)},anchor=north west,draw=none}]
\addplot[mark=*] coordinates {\SizeMappedCoords};
\addlegendentry{mapped}
\addplot[mark=square*,gray] coordinates {\SizeCopyCoords};
\addlegendentry{copy}
\end{axis}
\end{tikzpicture}
\end{minipage}
\caption{The public mapped path at scale: (a) the working-set sweep,
with the matched protocol's \MachRead\,GB/s pure-read reference dashed;
(b) dispatch granularity for the mapped and copy paths.  The
two panels use different vertical ranges.}
\label{fig:scale}
\end{figure*}

At scale the mapped path rises to \WsBigMap\,GB/s across
\WorkingSetPoints\ working-set points spanning
\WorkingSetMinGB--\WorkingSetMaxGB\,GB (Fig.~\ref{fig:scale}a), above
the matched protocol's \MachRead\,GB/s pure-read reference: a 16\,GB
pass amortizes the dispatch and synchronization boundaries of the
12-dispatch protocol, so that reference is the protocol's
measured ceiling rather than a machine limit.  Dispatch granularity governs how quickly that
rate is reached: in the matrix-size sweep the mapped path rises from
\SizeSmallMap\ to \SizeBigMap\,GB/s while the copy path plateaus near
\SizeCopyCap, dispatch cost dominates below roughly \SizeCrossMB\,MB,
and even at the smallest point mapping leads by \SizeSmallRatio$\times$
(Fig.~\ref{fig:scale}b).

The advantage also persists as the decode batch grows.  In a
$T$-sweep holding one multi-column kernel and both byte paths fixed
while the activation-column count $T$ rises from 1 to \TaxTMax\ (five
processes under the protocol above), adoption stays ahead at every
point, narrowing from \TaxTOne$\times$ at $T{=}1$ through \TaxTTwo,
\TaxTFour, and \TaxTEight$\times$ to \TaxTSixteen$\times$ at
$T{=}\TaxTMax$, as the aggregate logical weight rate rises under that
kernel from \TaxTokLo\ to \TaxTokHi\,GB/s.

\section{What each condition costs}\label{sec:conditions}

We vary the three conditions of Table~\ref{tab:contract} in one
factorial design with a shared Python host path, allocator,
\texttt{compile\_shader} kernel, and framework stream.  C1 toggles adopting versus re-ingesting
the mapped pages; C2, accelerator-resident activations versus a host round
trip; C3, GPU-side order versus a host-visible drain.  Cell order is
randomized independently in every process.  The estimates below apply to
this execution model.  The cells use two code paths
(\S\ref{sec:routes}): the
all-public path (standalone producer, public kernel surface) and an
internal extension used for the foreign-queue and ordering cells.

\subsection{Eight implementations at idle}
We first measure all eight implementations with the queue idle.  The cube varies mapped versus copied weights, device-resident versus
round-tripped activations, and framework-stream versus own-queue dispatch.
The Metal source and tensor shapes are identical; only storage and handoff
change.

\begin{table}[t]
\centering\small\setlength{\tabcolsep}{3.5pt}
\renewcommand{\arraystretch}{1.07}
\begin{tabular}{@{}lrr@{}}
\toprule
\multicolumn{3}{@{}l}{\emph{(a) idle queues, \SurveyProcesses\ processes}}\\[2pt]
Cell (weights, activations, queue) & GB/s & IQR \\
\midrule
map, device, stream & \CubeMapDevStream & \CubeMapDevStreamIQR \\
map, device, own & \CubeMapDevOwn & \CubeMapDevOwnIQR \\
map, round trip, own & \CubeMapRtOwn & \CubeMapRtOwnIQR \\
map, round trip, stream & \CubeMapRtStream & \CubeMapRtStreamIQR \\
copy, device, stream & \CubeCopyDevStream & \CubeCopyDevStreamIQR \\
copy, round trip, stream & \CubeCopyRtStream & \CubeCopyRtStreamIQR \\
copy, device, own & \CubeCopyDevOwn & \CubeCopyDevOwnIQR \\
copy, round trip, own & \CubeCopyRtOwn & \CubeCopyRtOwnIQR \\
\midrule
resident, same kernel & \CubeRefResident & \CubeRefResidentIQR \\
stock framework path & \CubeRefStock & \CubeRefStockIQR \\
\midrule
\multicolumn{3}{@{}l}{\emph{(b) $+$ in-flight framework load, \FactorialProcesses\ processes}}\\[2pt]
Cell & excess ms/pass & IQR \\
\midrule
C1+C2+C3 & \FactFull & \FactFullIQR \\
remove C3 & \FactCthreeOff & \FactCthreeOffIQR \\
remove C2 & \FactCtwoOff & \FactCtwoOffIQR \\
remove C1 & \FactConeOff & \FactConeOffIQR \\
remove all & \FactAllOff & \FactAllOffIQR \\
\bottomrule
\end{tabular}
\caption{The decomposition: (a) the idle cube, with two reference arms
below the rule; (b) excess milliseconds per
1.06\,GB pass over the in-flight framework workload alone.}
\label{tab:cube}
\end{table}

Table~\ref{tab:cube}(a) lists the eight cells together with two
reference arms: the same kernel over resident weights, and the stock
framework path.  Every copy cell
remains within \CubeCopyLo--\CubeCopyHi\,GB/s, and
own-queue mapping has the highest idle rate because it batches dispatches
and waits once, as a standalone runtime does.  During eager framework
execution, the same storage must instead be ordered against work already
in flight, and a host-visible exchange drains that work;
\S\ref{sec:live-reversal} measures the reversal this produces.

\subsection{Under load, removing any condition adds measurable cost}
Against an in-flight framework load (Table~\ref{tab:cube}(b)), the
full cell adds no measurable excess, and removing any one condition
adds a cost well outside the full cell's IQR.  The full $2^3$ model over
process-level paired contrasts across the 8 processes puts the
C1, C2, and C3 main effects at $+\FactEffCone$, $+\FactEffCtwo$, and
$+\FactEffCthree$\,ms, with every interaction spanning zero except
C2$\times$C3 (\FactEffCtwoCthree\,ms, \FactEffCtwoCthreeCI; the
C1$\times$C2, C1$\times$C3, and three-way intervals are
\FactEffConeCtwoCI, \FactEffConeCthreeCI, and
\FactEffThreeWayCI\,ms).  This
interaction reflects partial masking of ordering overhead by the host round
trip, which explains why C3's single-condition ablation
($+\FactNoCthree$\,ms relative to the full cell) exceeds its averaged main
effect.  The C1 effect is additive with the C2 and C3 effects.

The loaded-queue setting of Table~\ref{tab:cube}(b) also explains
why a per-dispatch rate can improve when more framework work is
present: fine-grained weight reads fill encoder gaps, and a dense
queue raises the operating clock.  GPU timestamps show serialized
encoder intervals rather than simultaneous kernels; queue occupancy is
the state variable missing from the idle cube.

In a depth sweep, as framework work in flight grows $8\times$,
the in-stream cell's excess stays flat; round-trip cells add roughly a
third of a millisecond per host exchange while moving a fixed 77\,KB, which
places the additional cost in the drain rather than in the transfer;
and the copy-cell penalty, which partially overlaps unrelated GPU
work, falls from 17.4 to 9.5\,ms.  The copy penalty is therefore
largest at shallow queue depth, while the round-trip cost is paid per
exchange regardless of the bytes moved.

\subsection{Ordering is about host visibility}
C3 permits both framework-stream ordering and GPU-side event ordering.  A
second randomized campaign compares stream order, a plain on-GPU event, and
a host drain.
With C1 and C2 enabled, the three cells cost \OrdStreamCell,
\OrdEventCell, and \OrdDrainCell\,ms.  The paired event-minus-stream interval
is \OrdEventCI; the experiment does not resolve those two GPU-side mechanisms.
The host drain is slower, with paired interval \OrdDrainCI.  The
operational requirement is therefore to establish dependencies on the
GPU; in-stream encoding is the fastest measured implementation.

\begin{figure}[tb]
\centering
\resizebox{\linewidth}{!}{%
\begin{tikzpicture}[font=\footnotesize, x=1mm, y=-1mm]
\definecolor{fw}{gray}{0.76}\definecolor{wt}{gray}{0.34}
\fill[fw] (0,-7.2) rectangle (5,-4.2);
\node[anchor=west] at (5.7,-5.7) {framework's own work};
\fill[wt] (38,-7.2) rectangle (43,-4.2);
\node[anchor=west] at (43.7,-5.7) {weight reads};
\node[anchor=west] at (63,-5.7) {excess (ms)};
\node[anchor=east] at (-1.5,2.6) {in-stream};
\fill[fw] (0,0.5) rectangle (16,4.7);
\fill[wt] (16,0.5) rectangle (28,4.7);
\fill[fw] (28,0.5) rectangle (44,4.7);
\node[anchor=west] at (63,2.6) {\OrdInStream};
\node[anchor=east] at (-1.5,17.1) {\texttt{MTLEvent}};
\fill[fw] (0,12.5) rectangle (16,16.7);
\fill[fw] (17,12.5) rectangle (33,16.7);
\fill[wt] (16,17.5) rectangle (28,21.7);
\node[anchor=west, gray] at (34.5,19.6) {\itshape queues overlap};
\node[anchor=west] at (63,17.1) {\OrdEvent};
\node[anchor=east] at (-1.5,31.6) {\texttt{MTLSharedEvent}};
\fill[fw] (0,29.5) rectangle (16,33.7);
\fill[wt] (23,34.5) rectangle (35,38.7);
\fill[fw] (39,29.5) rectangle (51,33.7);
\node[anchor=west, gray] at (17,27.0)
  {\itshape host-visible signal; overlap lost};
\draw[->, gray] (16,33.2) -- (22.6,35.6);
\draw[->, gray] (35,36.2) -- (38.6,33.4);
\node[anchor=west] at (63,31.6) {\OrdShared};
\node[anchor=east] at (-1.5,47.1) {host drain};
\node[anchor=west, gray] at (17,43.6) {\itshape every in-flight op retires};
\fill[fw] (0,45.0) rectangle (16,49.2);
\draw[pattern=north east lines, pattern color=gray, draw=gray]
  (16,45.0) rectangle (46,49.2);
\fill[wt] (46,45.0) rectangle (58,49.2);
\node[anchor=west] at (63,47.1) {\OrdDrain};
\node[anchor=west, gray] at (0,54.5)
  {\itshape one 1.06\,GB pass; time increases from left to right};
\end{tikzpicture}}
\caption{The ordering hierarchy; each excess is the median of process
medians across \OrdProcesses\ independent processes, 10 timed passes
each (process-level 95\% intervals: in-stream \OrdInStreamCI, event
\OrdEventLocalCI, shared \OrdSharedCI, drain \OrdDrainMechCI).  In-stream
work largely occupies an existing encode gap.  A plain
\texttt{MTLEvent} keeps the dependency on-GPU and the queues overlap.
\texttt{MTLSharedEvent}'s host-observable semantics coincide with lost
overlap; the experiment does not identify which internal component causes
the cost.  A host drain retires every in-flight operation first.  Its
measured cost increases with queue depth.  Each excess is the time
added over the in-flight framework workload alone, as in
Table~\ref{tab:cube}(b).  The event, shared-event, and drain rows are
driven by a standalone foreign-queue dispatcher; the campaign's
excesses, the in-stream row included, are not comparable with the
in-framework three-cell campaign in the text.  The timeline is schematic;
bar lengths are not proportional to the measured excess.}
\label{fig:ordering}
\end{figure}
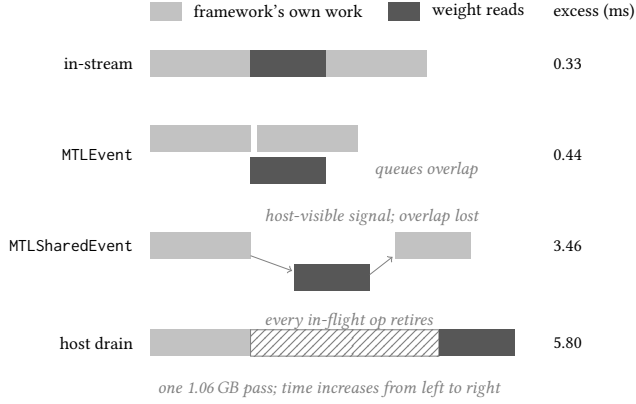

A separate campaign measures four ordering mechanisms for the same pass across
\OrdProcesses\ independent processes, with the foreign-queue cells driving
a standalone Metal dispatcher against the framework stream: in-stream
\OrdInStream\,ms, plain \texttt{MTLEvent} on the foreign queue \OrdEvent,
\texttt{MTLSharedEvent} \OrdShared, and host drain \OrdDrain\ (medians of
process medians; process-level intervals in Fig.~\ref{fig:ordering}).  The
median per-process paired difference between the foreign-queue event
and in-stream is \OrdEvMinusStream\,ms (interval
\OrdEvMinusStreamCI; the arm medians differ by 0.11\,ms), an order of
magnitude below any host-visible mechanism.  The large cost appears when
a signal becomes host-visible; changing queue ownership while retaining
GPU-side ordering adds much less.  With C2 off, the per-call round trip
has already serialized the pass, and the ordering mechanisms converge;
C3 has measurable overhead only when C2 is
enabled (the C2$\times$C3 term above).

\subsection{The live system reverses the idle ranking}
\label{sec:live-reversal}
The idle ranking of Table~\ref{tab:cube}(a) puts own-queue
mapping ahead of the framework-stream cell.
The live system reverses that conclusion: mapping on a foreign queue while
round-tripping activations takes \KthreeSpineForeign\,s, versus
\KthreeSpineStock\,s for stock ingestion.  This C1-only configuration
lacks C2 and C3.

A production-spine ablation applies the same removals to
\KthreeMatrices\ real matrices: the full configuration takes
\AblFull\,s; removing C3, C2, or C1 multiplies that time by
\AblNoCthree, \AblNoCtwo, and \AblNoCone\ (0.31, 0.52, and 1.96\,s),
and stock fp32 multiplies it by \AblStock\ (3.76\,s).  A separate
four-path experiment preserves the ordering of the copied and adopted
paths (Table~\ref{tab:k3-replay}); its copied arm reads a one-time
copy while the ablation's C1-removed cell re-ingests per use, so
absolute copy times are not compared across the two.  Removing dtype
inflation reduces the live and replay times by factors of 2.0
and 3.4; removing the copy yields further factors of 3.8 and
10.  Both experiments show that
the mapped storage must participate in the framework's activation and
ordering rules.

\begin{table}[tb]
\centering\footnotesize\setlength{\tabcolsep}{3.5pt}
\begin{tabular}{@{}lr@{}}
\toprule
Byte path & replay (s) \\
\midrule
stock fp32 ingestion & \ReplayStock \\
same kernel, copied int8 & \ReplayCopy \\
mapped + host round trip & \ReplayOwnRt \\
\textbf{adopted, in-stream} & \textbf{\ReplayMapped} \\
\bottomrule
\end{tabular}
\caption{A separate replay experiment applies the four byte paths of
Table~\ref{tab:k3} to the same \KthreeMatrices\ matrices.  It
preserves the ranking of the copied and adopted rows; absolute times
are not comparable with the live stage times of Table~\ref{tab:k3},
whose stage interval interleaves attention, norms, and sampling, and
the C1-only row is slower than stock only in the live
system, where it
synchronizes against in-flight framework work.  The ablation
experiment in the text measures \AblFull\,s for its full
configuration and 3.76\,s for stock fp32, within 5\% and 9\% of the
corresponding paths here (\ReplayMapped\ and \ReplayStock\,s).}
\label{tab:k3-replay}
\end{table}

C2 and C3 describe what ordinary framework code already does.  A tensor
produced by a framework operator stays in the accelerator-visible domain, and
work encoded into the current stream is ordered by the framework's own
dependency tracking.  An implementation loses both properties when it
leaves the framework.  The internal extension's foreign-queue path uses a
private queue and a host-observable completion; those are the two properties
whose cost the C1-only configuration above measures.  The contract therefore requires adoption
inside the framework execution path.  There, the last two conditions hold by
construction, which is why the all-public path of
\S\ref{sec:impl} satisfies the contract without loading the internal
extension.

\section{Implementation}\label{sec:impl}

\subsection{A framework-independent producer}\label{sec:producer}
The producer is a \ProducerLines-line C extension that compiles against
Metal, POSIX, and CPython only: no PyTorch or MLX headers, and no
linkage against either framework.  Each call opens one page-aligned
tensor file, maps it read-only with \textsc{map\_shared}, wraps the
range using \texttt{newBufferWithBytesNoCopy} on the system Metal
device, and returns a versioned \texttt{kDLMetal} capsule with the
read-only flag set and explicit compact strides \citep{dlpack}.
\texttt{torch.from\_dlpack} or \texttt{mx.from\_dlpack} then adopts the
capsule as ordinary framework storage; the MLX import runs without
PyTorch installed.

\begin{algorithm}[tb]
\caption{The file-backed tensor producer.}
\label{alg:producer}
\begin{algorithmic}[1]
\REQUIRE tensor file; shape $(M,K)$; dtype $\tau$
\STATE $n \gets \mathrm{size}(path)$
\IF{$n \neq M{\cdot}K{\cdot}\mathrm{sizeof}(\tau)$ \OR $n \bmod \mathrm{pagesize} \neq 0$}
  \STATE \textbf{reject}
\ENDIF
\STATE $p \gets \mathrm{mmap}(path,\ n,\ \textsc{prot\_read},\ \textsc{map\_shared})$
\STATE $b \gets \textsc{NewBufferWithBytesNoCopy}(p,\ n)$
\STATE close the descriptor
\STATE $ctx \gets \langle b,\ p,\ n\rangle$
\STATE $t \gets \langle data{=}b,\ \textsc{kDLMetal},\ \tau,\ shape{=}(M,K)\rangle$
\STATE $t.strides \gets (K,1)$;\quad $flags \gets \textsc{read\_only}$
\STATE $deleter \gets \lambda:\ \textsc{Release}(b);\ \mathrm{munmap}(p,n)$
\RETURN versioned capsule $\langle t,\ ctx,\ flags,\ deleter\rangle$
\end{algorithmic}
\end{algorithm}

The guards in Algorithm~\ref{alg:producer} encode the constraints
measured in \S\ref{sec:constraints}.  Line~2 rejects partial pages because Metal
wires whole pages and a partial page would alias a neighboring tensor;
line~10 uses explicit strides because DLPack $\ge$1.2 requires non-null
strides for tensors with nonzero rank \citep{dlpack}, and one tested
consumer faults on null strides.

Lifetime follows the consumer.  The capsule owns exactly one mapping and one
buffer; nothing is cached across calls.  If the capsule is consumed,
releasing the adopted tensor releases the buffer and unmaps the range;
if it is never consumed, destroying the capsule performs the same
cleanup.  The deleter touches no Python state and may run from any
thread.  Tests confirm content identity, file-write aliasing,
misalignment rejection, and tensor lifetime after all producer
references are released; 200 consumed and 200 unconsumed capsule
cycles leave the process footprint flat.

DLPack's read-only flag records producer intent, while enforcement remains
consumer-dependent \citep{dlpack}.  Both tested consumers import the
versioned capsule and still admit in-place stores.  Behavior under a mutating
kernel then depends on the substrate.  In our
probes, a GPU store to the read-only Metal mapping is silently
discarded, the CUDA path faults, and the tested Windows Vulkan drivers
reject the read-only view during import (\S\ref{sec:apu}).  A framework-level
mapped-storage interface should propagate immutability and reject
mutating operations before dispatch; DLPack import alone cannot
guarantee that behavior.

\subsection{Which route each measurement uses}\label{sec:routes}
For read-only downstream operations, the imported tensor participates as
ordinary MPS or MLX storage.  PyTorch's custom-kernel surface encodes
against it in the framework stream, supplying C2 and C3 without private
scheduler hooks.  This all-public path (standalone producer, public
import, public kernel surface) is the primary implementation.  The
matched adoption campaigns of \S\ref{sec:tax} and the model
campaigns of \S\ref{sec:eval} measure adoption through it, with two
exceptions: \KthreeName\ integrates through
the internal extension described next, and the first held-out session
(\S\ref{sec:heldout}) uses that extension's producer.  The
standalone producer is $\ProdRateDiffPct$\% faster than that
producer (\gbs{\ProdRateStd} versus \gbs{\ProdRateExt},
median paired per-pass difference), and GEMV outputs are bitwise
identical.

The multi-token kernel, the ordering study, and the \KthreeName\
production integration use an
internal extension; a
spine-specialized variant runs the production-spine experiment, and a
multi-column variant runs the $T$-sweep of \S\ref{sec:scale}.  For these mechanism studies the extension caches mappings by path for
the life of the process; the deployable path instead uses the
per-capsule lifetime of \S\ref{sec:producer}.  It encodes into PyTorch's current Metal encoder on the stream's serial
dispatch queue.  It does not end the encoder, allowing the framework to
coalesce surrounding work.  Its packed-int8 GEMV serves up to
\KernelMaxT\ activation columns so speculative
batches~\citep{speculative} reuse each weight read, and reaches \ResOurs\,GB/s on resident weights; adoption changes storage ownership while leaving the arithmetic
unchanged.

\subsection{Constraints that determine correctness}\label{sec:constraints}
At model scale, correctness depends on shared mappings, page-aligned
imports, and framework-serialized command encoding.

First, the mapping must remain shared.  A private mapping is converted
to anonymous pages when Metal wires it, recreating the copy at model
scale: a \PrivMapGB\,GB private mapping produced \PrivAnonGB\,GB of
anonymous memory, $\PrivWireSlowdown\times$ slower wiring, and a
$\PrivTputLoss\times$ throughput loss.  This behavior was observed on
macOS; Linux and Windows keep private file mappings file-backed.
Small tests can return correct values even when a large wired private
mapping would become an anonymous copy.

Second, the imported base and length must satisfy Metal's page
constraints.  Per-tensor files make this explicit, and an aligned
container format would provide the same property without requiring
one inode per
tensor; container checkpoints otherwise need aligned tensor offsets or
a one-time relayout into per-tensor files.  Our converter performs
that relayout once; adoption then maps the \KthreeFiles-file spine
(one weight file and one scale file per matrix) in \KthreeMapOnly\,s
and wires it in \KthreeWireOnly\,s at \KthreeWireGBs\,GB/s,
\KthreeFirstTouch\,s in total.

Third, command encoding must follow the framework's serialization rule:
the internal path enters PyTorch's stream dispatch queue before touching
the current encoder and leaves the encoder open.  Encoding from an
arbitrary host thread can race the framework and terminate the process;
ending the encoder per weight destroys the batching that makes C3
effective.  The public DLPack path does not encounter this constraint
because the framework is the sole encoder.

\section{Evaluation}\label{sec:eval}

The evaluation asks four research questions.  \textbf{RQ1}: does
adoption remove the tax end to end while matching the throughput of resident
storage?  \textbf{RQ2}: what does adoption require of model
code, and what does it change at capacity and startup?  \textbf{RQ3}:
how does adopted storage behave across processes and under memory
pressure?  \textbf{RQ4}: do
the mechanism and the endpoint procedure transfer across architectures
and to a held-out machine (\S\ref{sec:arch}, \S\ref{sec:heldout})?

\subsection{Protocol controls and alternative explanations}\label{sec:controls}
Table~\ref{tab:controls} summarizes each protocol control and the
measurement it affects.  Apple GPU frequency ramps over several
passes, unequally across mechanisms, so every Apple rate campaign discards
\WarmupPasses\ passes and rotates arm order; timing the ramp can
understate the mapped arm several-fold.  In the five-process
matrix-size sweep (Fig.~\ref{fig:scale}b), the mapped path's first
timed pass reaches
2--84\% of its steady rate while the clock ramps, and a single-discard
estimator reports 134--167\,GB/s at 16.8\,MB where the twelve-discard
median is 328\,GB/s.  The
copy path stays on the copy engine and does not ramp the same way, so
insufficient warm-up biases the measured gap between the arms.

This warm-up behavior is specific to the tested
Apple systems.  In an A10 campaign across \AudProcs\ independent
processes, every CUDA cell reaches steady state by pass
\AudSteadyPass, and applying the Apple discard changes every CUDA result by
less than \AudCorrection\%; the CUDA and GH200 campaigns therefore use
a single warm-up.  A rotated
same-process session (mapped \PairSessMapped, resident
\PairSessResident, pure read \PairSessPure\,GB/s) agrees with the
independent-process result of \S\ref{sec:pairedcamp}, so session drift
does not account for the mapped-minus-resident difference reported
there.

The \emph{endpoint} a probe measures---map, resident, or per-use
copy (\S\ref{sec:design})---must also match the deployed byte path.
A probe that recycles a resident buffer smaller than the
working set measures cache residency instead, and would select a
different endpoint under the procedure of
\S\ref{sec:design} (\S\ref{sec:apu}).

\begin{table}[tb]
\centering\small\setlength{\tabcolsep}{4pt}
\renewcommand{\arraystretch}{1.08}
\begin{tabular}{@{}ll@{}}
\toprule
Threat & Control \emph{(measurement affected)} \\
\midrule
clock ramp & discard \WarmupPasses\ passes, rotate arms \emph{(rates)} \\
cache-sized endpoint & hold the full deployed set \emph{(policy)} \\
kernel confound & one mapped/resident GEMV \emph{(attribution)} \\
checkpoint format & count weight bytes read per token \emph{(rates)} \\
page-cache eviction & residency queried before/after \emph{(regime)} \\
unevaluated lazy graph & force output evaluation \emph{(MLX rates)} \\
\bottomrule
\end{tabular}
\caption{Controls addressing alternative explanations.}
\label{tab:controls}
\end{table}

Mutation visibility establishes
storage aliasing; checksums verify that the kernel reads the intended mapping
and that eviction preserves bytes; and end-to-end runs compare logits, greedy
tokens, or fixed-oracle outputs.  Lazy execution is forced at the
measurement boundary: an unevaluated MLX result reports rates above
physical bandwidth, so every MLX arm evaluates its outputs before
synchronization.

Matched comparisons share prompt, token count, format, kernel, process
count, and arm rotation; native MLX and llama.cpp runs use different
formats or protocols, so their rates are not directly comparable with
the matched arms and are reported as references.  At the evaluated
shape, PyTorch's
\texttt{w\,@\,x} form reaches \gbs{60} against \gbs{258} for
\texttt{F.linear} over the same already-resident
tensors---arithmetic-form rates with no ingestion---so the stock arms
use \texttt{F.linear}.

\begin{table}[b]
\centering\small\setlength{\tabcolsep}{2pt}
\begin{tabular}{@{}lrrc@{}}
\toprule
Spine path & stage (s) & token (s) & contract \\
\midrule
stock fp32 ingestion & \KthreeSpineStock & \KthreeTokenBefore & C2+C3 \\
same kernel, copied int8 & \KthreeSpineCopy & n/a & C2+C3 \\
mapped + host round trip & \KthreeSpineForeign & n/a & C1 only \\
\textbf{adopted, in-stream} & \textbf{\KthreeSpineMapped} &
  \textbf{\KthreeTokenAfter} & \textbf{C1+C2+C3} \\
\bottomrule
\end{tabular}
\caption{The \KthreeName\ spine, live (\KthreeLiveRuns\ runs).  The
contract column lists the conditions each path satisfies; stock
ingestion runs inside the framework (C2+C3) but re-creates the copy C1
removes.  \emph{n/a} marks paths measured at stage granularity only;
full-token latency was recorded for the two deployed configurations.
K3 cells run on the internal extension (\S\ref{sec:routes}).}
\label{tab:k3}
\end{table}

\subsection{RQ1: the motivating system and a matched dense model}
On \KthreeName, the spine stage over adopted storage falls from
\KthreeSpineStock\ to \KthreeSpineMapped\,s; per-token latency falls from
\KthreeTokenBefore\ to \KthreeTokenAfter\,s
(\KthreeTokenSpeedup$\times$), the remainder being routed-expert storage
traffic this work leaves unchanged (Table~\ref{tab:k3}).  Packing alone
reaches \KthreeSpineCopy\,s; holding the packed kernel fixed,
adoption reduces the stage to \KthreeSpineMapped\,s, a further
\KthreeAdoptOnly$\times$.  With fixed
routing and oracle tokens, the compared byte paths produce
byte-identical outputs.

\begin{table}[tb]
\centering\small\setlength{\tabcolsep}{3pt}
\begin{tabular}{@{}lrrl@{}}
\toprule
\QwenModel\ int8 & tok/s & logical GB/s & storage \\
\midrule
\textbf{public adoption} & \textbf{\QwenPublic} & \textbf{\QwenPublicRate} & \textbf{page cache} \\
load-once resident & \QwenResident & \QwenResidentRate & owned copy \\
internal extension & \QwenExt & \QwenExtRate & page cache \\
stock per-use ingestion & \QwenStock & \QwenStockRate & transient copy \\
\bottomrule
\end{tabular}
\caption{Matched \QwenModel\ runs, \QwenProcesses\ processes per arm
(resident $n{=}\QwenResProcs$).  \emph{Logical GB/s} is tok/s times the
\QwenBytes\,GB of weight bytes the model reads per token.  Bold marks
the public adoption path, not the fastest arm.}
\label{tab:qwen}
\end{table}

On a matched dense model, the public route reaches \QwenPublic\,tok/s
with interval [\QwenPublicLo, \QwenPublicHi], versus \QwenStock\,tok/s
with interval \QwenStockCI\ (Table~\ref{tab:qwen}; logical rate is
\QwenBytes\,GB per token times tok/s---the model reads its full weight
set per token, so the per-token byte count equals the checkpoint
footprint).  A load-once resident control (the identical kernel over a
one-time owned copy of the same int8 bytes) reaches
\QwenResident\,tok/s, with the adopted median within 1.3\% of it.  The
end-to-end arms are unpaired ($n{=}\QwenProcesses$ and
$n{=}\QwenResProcs$); throughput parity is therefore assessed with the
paired microbenchmark of \S\ref{sec:pairedcamp}, and the end-to-end
values are descriptive.  The two arms differ in
what persists after initialization: \QwenBytes\,GB of reclaimable file
pages versus a second owned copy.  The internal extension reaches
\QwenExt\,tok/s with a wider [\QwenExtLo, \QwenExtHi] interval from
per-layer host work.

\finding{RQ1}{Adoption removes the tax without reducing kernel throughput:
\AdoptPt\,GB/s on the public route, mapped-minus-resident bounded in
\PairDiffCI\ (paired).  On \KthreeName\ the adopted spine stage is
\KthreeStageSpeedup$\times$ faster (\KthreeAdoptOnly$\times$ from
storage alone); on a matched \QwenModel, adoption is
\QwenStockRatio$\times$ faster than per-use ingestion and within 1.3\%
of a load-once resident control.}

\subsection{RQ2: stock modules, capacity, and startup}
Storage adoption does not require a bespoke model implementation.  Rebinding
each two-dimensional parameter of a stock \texttt{transformers} model
through \texttt{load\_state\_dict} with \texttt{assign=True} preserves its
modules, \texttt{F.linear}, and \texttt{generate}.  On \CompatibilityModel,
across \CompatPairs\ process rounds per arm with arm order alternating, the
last-position logits over a fixed probe sequence are bitwise identical
between the mapped and stock arms and across every run, and decode
rates match (\CompatMappedTps\ versus \CompatStockTps\,tok/s).  An
independent sharing experiment measures the same two configurations at
22.9 and 23.0\,tok/s with the order reversed
(Table~\ref{tab:sharing}, where the stock loader appears as the
\emph{resident} arm); the difference is run-to-run variation.  The process's
peak resident set is \CompatMappedRSS\,GB against \CompatStockRSS\,GB:
the weight bytes still occupy DRAM, but as shared, reclaimable page-cache
pages rather than process-owned allocations.

At capacity, an unmodified \CapacityModel\ checkpoint occupies
\CapacityCheckpointGB\,GB of the \ReferenceRamGB\,GB machine.  A
campaign of \CapRounds\ rounds runs three arms, one process each, order
rotated: the stock loader; a rebind arm over page-aligned per-tensor
files; and a copy control sharing the rebind arm's files and
construction but materializing resident copies.  All three decode at \CapTok\,tok/s with
bitwise-identical probe logits; they differ before the first token.
Stock spends \CapStockParse\,s parsing shards and constructing modules,
and \CapStockCopy\,s copying: first token at \CapStockTTFT\,s.
Rebinding reaches it at \CapRebindTTFT\,s.  Across rounds, the median
paired TTFT ratio is \CapLoadRatio$\times$ with interval
\CapLoadRatioCI; the ratio of the arm medians is 5.4$\times$.  The map
itself takes \CapRebindLoad\,s, and the remaining \CapRebindFirst\,s
of first-token latency is dominated by page-fault wiring.  The copy control reaches
\CapCopyTTFT\,s: most of the gap against stock comes from the relayout,
which removes shard parsing, and the storage change contributes a median
per-round paired ratio of \CapAdoptOnlyStartup$\times$ on identical
files (arm medians \CapCopyTTFT\ versus \CapRebindTTFT\,s, a
1.21$\times$ ratio).  The
weights persist afterwards as reclaimable file pages.  A cold-state variant (\CapColdRounds\
rounds, the page cache evicted before every arm) yields the same
ordering: \CapColdStockTTFT/\CapColdRebindTTFT/\CapColdCopyTTFT\,s,
per-round ratio \CapColdLoadRatio$\times$ (arm medians give
4.7$\times$).  The warm and cold campaigns ran separately, so their
absolute times are compared only within each campaign.  The
production spine likewise pays its mapping and wiring costs only at
startup (\S\ref{sec:constraints}); later tokens reuse the mappings.

\finding{RQ2}{Adoption needs no bespoke model code: rebinding stock
\texttt{transformers} modules leaves probe logits bitwise identical and
decode rates unchanged.  What changes is startup and residency: a
\CapacityCheckpointGB\,GB checkpoint cuts time to first token by
\CapLoadRatio$\times$ (\CapAdoptOnlyStartup$\times$ from storage
alone, paired per round on identical files), and the weights stay
reclaimable afterwards.}

\subsection{RQ3: sharing and page-cache behavior under pressure}
Adoption changes more than the steady-state rate: clean mapped pages
are reclaimable without write-back, while an anonymous framework copy
must be compressed, swapped, or retained.  We track the distinction with
\texttt{mincore} bitmaps, VM counters, and GPU checksums after
eviction.  Under pressure from an idle anonymous competitor, the OS
takes memory from elsewhere---compressing or reclaiming other
allocations---and preserves the mapping
touched on every pass.  With a 25\,GiB
mapped set and the idle competitor---an incompressible anonymous
region---grown to 105\,GB on the
128\,GB machine, the mapped set remains 100\% resident and
\texttt{vm\_stat} pageouts stay at zero: the compressor absorbs the
machine's other anonymous memory, and no clean mapped pages are
written back or dropped.  A private mapping
under the same sweep falls to 71\% residency, because its privatized
anonymous copy competes for the same memory.  When the competitor
instead touches its pages continuously, the two
sets evict one another once their sum passes
\HotSetLo--\HotSetHi\% of installed memory on the tested machine, and
alternating passes fall to storage speed.  Checksums after
complete eviction reproduce the file values (a fully evicted 1\,GiB
file matches ground truth after 73\,ms of refault I/O), confirming
that reclaim and refault preserve the data.

The representation-count difference increases with the number of
processes: mapped decoders share one set of file-backed pages and add
essentially no anonymous memory beyond the first process's load, while
resident loaders add anonymous memory with every process until, at
capacity, the duplicates push the machine into compression, evict
\ShareBigFileEvict\,GB of the shared cache, and drop decode by two
orders of magnitude (Table~\ref{tab:sharing}).

\begin{table}[tb]
\centering\footnotesize\setlength{\tabcolsep}{4pt}
\begin{tabular}{@{}llrrr@{}}
\toprule
model, $N$ & arm & tok/s per process & $\Delta$file (GB) & $\Delta$anon (GB) \\
\midrule
7B, 1 & mapped & 22.9 & +15.4 & +0.4 \\
7B, 1 & resident & 23.0 & +15.2 & +11.0 \\
7B, 2 & mapped & 18.1, 18.8 & 0.0 & +0.8 \\
7B, 2 & resident & 16.9, 17.1 & 0.0 & +11.8 \\
7B, 4 & mapped & 15.3--20.1 & 0.0 & +1.1 \\
7B, 4 & resident & 12.5--16.8 & 0.0 & +32.6 \\
32B, 2 & mapped & \ShareBigMapTokHi, \ShareBigMapTokLo & +\ShareBigFile & +1.6 \\
32B, 2 & resident & \ShareBigResTok, 0.07 & $-$\ShareBigFileEvict & +\ShareBigAnon \\
\bottomrule
\end{tabular}
\caption{Sharing across processes: $N$ concurrent decoders of one
checkpoint, with \texttt{vm\_stat} deltas sampled against a settled
baseline.  Mapped arms add file-backed bytes only when the checkpoint
first enters the cache and essentially no anonymous memory afterwards;
resident arms add anonymous memory with every process until, at
capacity, the duplicates evict the shared cache.  The 7B rows run
\CompatibilityModel; the 32B rows run the \CapacityModel\ checkpoint
of the capacity campaign.  Outputs are
byte-identical in every cell.}
\label{tab:sharing}
\end{table}

\finding{RQ3}{Adopted weights remain clean, reclaimable page-cache pages:
checksums survive eviction and refault, and $N$ processes share one
mapped copy while
resident loading creates $N$ copies (at capacity, \ShareBigMapTokHi\
versus \ShareBigResTok\,tok/s).}

\section{Architecture boundaries}\label{sec:arch}

File-backed adoption is specific to memory topology and should be selected
per platform.  We exercise the same choice on an integrated Vulkan GPU, a
coherent CPU--GPU link, and a discrete PCIe GPU, and survey a rented
five-machine CUDA fleet that places the endpoints on a common scale
(Table~\ref{tab:fleet}).  Together with the
Apple-silicon systems of \S\S\ref{sec:tax}--\ref{sec:eval}, these
platforms separate software ownership from physical placement: on Apple silicon
and the APU, mapped and framework-owned storage share one DRAM domain,
so a second allocation duplicates bytes without a faster tier; on
GH200, once the model exceeds HBM, the live comparison is mapping
against link-bound streaming, both on the coherent link; on the
discrete GPU every mapped read traverses PCIe, while a one-time copy
moves the weights to a materially faster placement.

\subsection{Integrated Vulkan deployment}\label{sec:apu}
On an \ApuPlatform, llama.cpp's Vulkan backend reports unified memory yet
its stock loader retains both the mapped model and a Vulkan allocation in
one DRAM pool.  We connect the loader's existing \texttt{buffer\_from\_host\_ptr} hook
to \texttt{VK\_EXT\_\allowbreak external\_\allowbreak
memory\_\allowbreak host}, advertised only on unified-memory devices;
the loader and the inference loop are unchanged.

The driver caps a single imported allocation at 2\,GiB, so the \LcppModelGB\,GiB
model is split into \LcppShards\ files with llama.cpp's own tool.
Across \ApuRuns\ paired processes, adoption raises decode from
\LcppStockTps\ to \LcppAdoptTps\,tok/s, a $\LcppSpeedup\times$
improvement, while peak working set falls from \LcppStockWs\ to
\LcppAdoptWs\,GiB (Table~\ref{tab:apu-system}).
Relative to the model file, those footprints are \LcppStockRatio$\times$ and
\LcppAdoptRatio$\times$.  Greedy
generation is byte-identical.  The platform also exposes a cumulative
importable-host-memory ceiling of \ApuHeapCeil\,GiB; sharding satisfies
the per-allocation limit but leaves that capacity boundary unchanged.

The footprint difference becomes latency under pressure: with
\CapPressFree\,GiB free and an \CapPressModel\,GiB model, stock peaks at
\CapPressStockWs\,GiB and completes in \CapPressStockWall\,s; adoption
peaks at \CapPressAdoptWs\,GiB and completes in \CapPressAdoptWall\,s.
The difference is the cost of carrying a duplicate inside one DRAM pool.

\begin{table}[tb]
\centering\small\setlength{\tabcolsep}{3pt}
\begin{tabular}{@{}lrr@{}}
\toprule
APU metric & stock / resident & adopted / mapped \\
\midrule
paired decode (tok/s) & \LcppStockTps & \LcppAdoptTps \\
peak working set (GiB) & \LcppStockWs & \LcppAdoptWs \\
pressure run wall time (s) & \CapPressStockWall & \CapPressAdoptWall \\
pressure-run peak set (GiB) & \CapPressStockWs & \CapPressAdoptWs \\
deployment endpoint (GB/s) & \ApuResDep & \ApuMapDep \\
\bottomrule
\end{tabular}
\caption{The integrated-GPU result at loader and system level.
Rows 1--2: the \LcppModelGB\,GiB three-shard model over \ApuRuns\
paired processes; rows 3--4: an \CapPressModel\,GiB model with
\CapPressFree\,GiB free; row 5: the full-working-set endpoint probe over the same deployed set
(\ApuW\,GB $=$ \LcppModelGB\,GiB).}
\label{tab:apu-system}
\end{table}

A deployment-scale endpoint probe explains the result: mapping reaches
\ApuMapDep\,GB/s, a whole-working-set resident buffer \ApuResDep\,GB/s,
and per-use copy \ApuCopyDep\,GB/s, so mapping leads the resident
endpoint by $\ApuRatioDep\times$.  A small-buffer probe
overstates the resident arm by $\ApuStageInflation\times$
(\ApuResCached\ versus \ApuResFullSet\,GB/s; \S\ref{sec:controls})
because it
repeatedly reads a cache-sized working set instead of holding the
deployed model.  Holding the byte total (\GranBytes\,MiB) and dispatch
shape fixed, a single import (through llama.cpp's buffer-size
override, which bypasses the advertised cap) and \LcppShards\ imports
produce the same
map-to-resident ratio (\ApuGranOne\ and \ApuGranThree), whereas
forcing the whole model through one oversized buffer makes adoption
14\% slower than stock in that paired configuration (2.94
versus 3.40\,tok/s): the reversal tracks the single import's size,
not import granularity.  Sharding changes feasibility; the
full-working-set endpoint probe determines performance.

The integration also shows the mechanism is not Metal-specific.
llama.cpp's Vulkan backend already declares the same loader-level
\texttt{buffer\_from\_host\_ptr} hook that its Metal backend
implements, but reports it unsupported on the tested device;
wiring that hook to \texttt{VK\_EXT\_\allowbreak external\_\allowbreak
memory\_\allowbreak host} is the only backend change; the hook
widens each imported host range to the driver's
alignment while keeping the byte offset, because rejecting unaligned
bases would
make ordinary checkpoint tensors ineligible for direct import.  With
alignment handled inside the hook, the loader, graph, and decode loop
use the
file pages directly.  Outputs are identical, and the footprint drop
reported above (Table~\ref{tab:apu-system}) is the removed Vulkan
duplicate, so the rate change follows from removing it.

The tested Windows
drivers reject a read-only file view at \texttt{vkAllocateMemory}, so
the loader imports a copy-on-write view instead, whose pages remain
file-backed and shared for a reader that never writes.  A standalone
Vulkan probe confirms in-place reading outside llama.cpp:
it imports one mapped weight file over a 1\,GiB working set and reads
it from a compute
shader at \ApuMapped\,GB/s---a probe-local rate, not the deployed-set
endpoint of Table~\ref{tab:apu-system}---while the process working
set stays at 1072\,MiB through import.

\subsection{Coherent link: capacity without a local copy}\label{sec:gh200}
\GhPlatform\ combines \GhHbmGB\,GB of HBM with \GhCpuGB\,GB of CPU
memory over a coherent link.  Table~\ref{tab:gh-endpoints} lists the
endpoints by physical route; the routes fall into three placement
classes---local HBM, link-bound traffic (bulk copy, overlapped
streaming, and direct Grace reads), and fault-driven migration.
File-backed mappings and pinned Grace allocations fall in the
direct-read subset of the link-bound class; ordinary CPU-first-touch
allocation does not.  Placement can be controlled while retaining file
semantics: in a single-file placement probe, the same file read from a
tmpfs bound to the HBM node reaches 3.0--3.1\,TB/s, while Grace-bound
tmpfs and ext4 read at 358--363 and 338--342\,GB/s (the
\GhMmap\,GB/s endpoint below is the full \GhBigGB\,GB set at
deployed import granularity).  The relevant distinction is placement
class; the labels ``CPU memory'' and ``GPU memory'' do not capture
this behavior.
For a \GhBigGB\,GB working set that exceeds HBM, mapping reads at
\GhMmap\,GB/s---93\% of the bulk-copy link rate---and overlapped
streaming reaches
\GhOverlap\,GB/s, both link-bound and inside the \PolTieBand\% band
(Table~\ref{tab:gh-endpoints}); mapping avoids a second staging
footprint.

\begin{table}[b]
\centering\small\setlength{\tabcolsep}{3pt}
\begin{tabular}{@{}lrp{0.32\columnwidth}@{}}
\toprule
GH200 endpoint & GB/s & physical route \\
\midrule
resident GEMV & \GhResident & local HBM \\
bulk copy & \GhLink & C2C link \\
overlapped streaming & \GhOverlap & C2C link \\
file-backed mapping & \GhMmap & direct Grace read \\
pinned host allocation & \GhHostAlloc & direct Grace read \\
plain host allocation & \GhMallocLo--\GhMallocHi & migration path \\
\bottomrule
\end{tabular}
\caption{GH200 endpoints under the paper's read protocol (medians of
5 runs, $\pm$2\%).  The resident row runs within HBM; the mapped row
reads the \GhBigGB\,GB working set; the plain-host-allocation row
falls in an active migration transient.}
\label{tab:gh-endpoints}
\end{table}

End to end, a stock
\texttt{transformers} model rebound to DLPack views of Grace-resident
file pages runs a \QwenFpSixteenGB\,GB fp16 \QwenModel\ at
\GhSeventyTwo\,tok/s despite exceeding HBM (mapping takes 2\,s, and
the 14\,s first token wires all 145\,GB), while anonymous
CPU-first-touch allocations enter a migration path: the end-to-end run
sustains only
\GhMigLo--\GhMigHi\,GB/s of weight reads, against the
\GhMallocLo--\GhMallocHi\,GB/s endpoint probe of
Table~\ref{tab:gh-endpoints}.  In first-touch probes, migration
reaches an HBM-resident steady state
when the set fits HBM (a 24\,GB set settles after 26--38\,s), while a
separate 109\,GB first-touch set reads at 70--140\,GB/s
mid-migration and does not settle.  The streaming regime considered
here exceeds
fast-memory capacity, so that steady state is unavailable.

On GH200 the mapped path shows no settling phase through the largest
tested set, while
anonymous CPU-first-touch allocation is still migrating; with a cold
cache the mapped path falls to storage rather than link speed, so
page-cache residency remains a precondition.

\subsection{Discrete PCIe: reject adoption}\label{sec:discrete}
Across the rented CUDA fleet, mapped reads track each machine's host
interconnect and resident reads track its device memory
(Table~\ref{tab:fleet}); the measured endpoint therefore follows
the memory topology.  The \texttt{malloc} column varies because the
allocation reaches the GPU through different mechanisms on different
platforms: implicit ingestion into
VRAM on the PCIe parts---on the A10, the \texttt{malloc} cell consumes
\AudMallocVram\,GB of VRAM while reading 2.11\,GB of matrices, whereas
the mapped-file cell consumes none---and a low-rate migration
transient on GH200.

\begin{table}[tb]
\centering\footnotesize\setlength{\tabcolsep}{3pt}
\begin{tabular}{lrrrr}
\toprule
 & \multicolumn{4}{c}{GB/s} \\
platform & resident & bus & mmap in place & \texttt{malloc} \\
\midrule
A10 & 470 & 25 & 25 & 400 \\
A100 SXM4 & 1268 & 26 & 27 & 846 \\
H100 PCIe & 1822 & 55 & 36 & 1779 \\
H100 SXM5 & 2859 & 54 & 35 & 2709 \\
GH200 (C2C) & \GhResident & \GhLink & \GhMmap & \GhMallocLo--\GhMallocHi \\
\bottomrule
\end{tabular}
\caption{GB/s across the rented fleet, five-run medians per cell.
File-backed in-place reads reach 65--104\% of each platform's
bulk-copy link rate and stay well below resident bandwidth.  The A10
warm-up campaign of \S\ref{sec:controls} (\AudProcs\ processes)
reproduces that machine's resident, mmap-in-place, and \texttt{malloc}
cells at \AudResident\,/\,\AudMmap\,/\,\AudMalloc\,GB/s.}
\label{tab:fleet}
\end{table}

On an \DiscretePlatform, a mapped host read reaches \DiscMapped\,GB/s
against an \DiscResident\,GB/s resident kernel.  Streamed weights
reach 51--57\,GB/s under the same-shape protocol, and three stacks
agree on the resident rate (torch 815.4, cupy 814.6, Vulkan
801.7\,GB/s).  Forcing the same
llama.cpp adoption path across PCIe lowers decode from \LcppDiscStock\
to \LcppDiscAdopt\,tok/s ($\LcppDiscSlowdown\times$).  Here a one-time
copy moves the weights to a faster placement and should be retained.  The
implementation, format, and loop match the APU experiment, while the memory
topology changes the outcome.

The API surface can also obscure an available direct-read path.  On an A10,
\texttt{cudaHostRegister} rejects a file-backed mapping under every tested
flag, with the same result for read-only and read-write mappings, while
anonymous memory and \texttt{malloc} storage register in the same
process.  \texttt{cudaHostRegisterReadOnly} returns \emph{operation not
supported} for every tested memory type.  The unregistered file mapping
can nevertheless be read in place by a kernel at bus rate
(Table~\ref{tab:fleet}).  Registration support and direct-read
capability must therefore be tested separately.

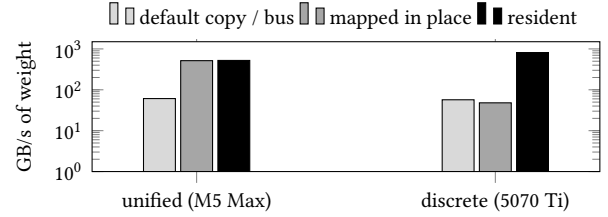
\begin{figure}[tb]
\centering
\begin{tikzpicture}
\begin{axis}[ybar, width=0.98\linewidth, height=3.3cm,
  ymin=1, ymax=1500, ymode=log, ylabel={GB/s of weight},
  symbolic x coords={unified (M5 Max),discrete (5070 Ti)},
  xtick=data, bar width=12pt, enlarge x limits=0.35,
  legend style={font=\footnotesize, at={(0.5,1.02)}, anchor=south,
                legend columns=3, draw=none},
  tick label style={font=\footnotesize}, label style={font=\footnotesize}]
\addplot[fill=gray!30] coordinates
  {(unified (M5 Max),\PtDefaultLo) (discrete (5070 Ti),\DiscBus)};
\addlegendentry{default copy / bus}
\addplot[fill=gray!70] coordinates
  {(unified (M5 Max),\AdoptPt) (discrete (5070 Ti),\DiscMapped)};
\addlegendentry{mapped in place}
\addplot[fill=black] coordinates
  {(unified (M5 Max),\CubeRefResident) (discrete (5070 Ti),\DiscResident)};
\addlegendentry{resident}
\end{axis}
\end{tikzpicture}
\caption{The same three byte strategies on both memory architectures (log
scale).  On discrete hardware, the mapped rate is limited by the
interconnect (\DiscMapped\ versus \gbs{\DiscResident}), so staging into VRAM is
the appropriate policy.  On unified memory, mapping reaches resident-class
bandwidth, while PyTorch's \texttt{.to("mps")} copy path
(\PtDefaultLo\,GB/s, the low end of the \PtDefaultLo--\PtDefaultHi\ range;
MLX constructor \MlxDefault) runs near the discrete bus rate
(\gbs{\DiscBus}), although no bus transfer is required.}
\label{fig:mirror}
\end{figure}

Figure~\ref{fig:mirror} compares the two architectures: the same
three byte strategies invert in ranking with the memory topology.  A
deployable interface
should therefore combine capability detection with a measured placement
policy; unconditional zero-copy selection is insufficient.

\section{Selecting the byte path}\label{sec:design}

\subsection{Choose among three endpoint paths}
At deployment, a weight byte has three relevant endpoints:
\emph{map}, which reads file pages in place; \emph{resident}, which copies
once into device-local storage and amortizes that copy; and
\emph{per-use copy}, which stages the bytes per use.  For working-set size
$W$ and achieved endpoint rate $B$, the measured path time is $W/B$.
Resident is excluded when the working set does not fit.  When two point
estimates differ by at most \PolTieBand\%, we prefer mapping because it
retains one evictable file-backed representation rather than a second owned
copy.  The band breaks ties in endpoint selection; it is not a
statistical equivalence criterion.  A sensitivity sweep from
\PolThetaLo\% to \PolThetaHi\% changes no verdict in
Table~\ref{tab:policy}; only below the GH200 pair's own \GhTieGapPct\%
gap does that tie resolve to per-use copy, and for no tested $\theta$
does the tie rule select mapping over a resident endpoint that leads
it by more than $\theta$.

Algorithm~\ref{alg:endpoint} defines the endpoint-selection procedure
evaluated in Table~\ref{tab:policy}.  We derive the rule from the four
measured platforms and evaluate it on one held-out machine.  The storage-bound stop of line~3 marks
the regime boundary; the guard of lines~7--8 requires post-execution
coexistence because accelerator allocation can be deferred until work
runs; and the tie rule prefers the one evictable representation.  A
resident arm smaller than $W$ measures cache residency instead, which
inflates one platform's endpoint by $\ApuStageInflation\times$
(\S\ref{sec:controls}).

\begin{algorithm}[tb]
\caption{Endpoint selection at deployment.}
\label{alg:endpoint}
\begin{algorithmic}[1]
\REQUIRE working set $W$: its files, sharding, kernel; band $\theta=\PolTieBand\%$
\STATE map $W$ at deployed import granularity; prefault
\IF{the mapped set is not resident}
  \STATE report storage-bound; \textbf{stop}
\ENDIF
\STATE $B_{map} \gets$ rate over the full mapped set
\STATE build a resident arm holding \emph{all} of $W$
\IF{allocation fails \OR the mapping is evicted after one pass of every arm}
  \STATE drop the resident endpoint
\ENDIF
\STATE $B_{dev}, B_{copy} \gets$ rates with the same kernel and set
\STATE drop endpoints that violate import or capacity limits
\STATE rank by $W/B$;\quad \textbf{if} leaders lie within $\theta$, choose map
\RETURN the surviving minimum
\end{algorithmic}
\end{algorithm}

\begin{table}[b]
\centering\small\setlength{\tabcolsep}{3pt}
\renewcommand{\arraystretch}{1.08}
\begin{tabular}{@{}lrrrll@{}}
\toprule
 & \multicolumn{3}{c}{$W/B$ (ms)} & \multicolumn{2}{c}{decision} \\
\cmidrule(lr){2-4}\cmidrule(l){5-6}
platform & map & res. & copy & rule & outcome \\
\midrule
M5~Max spine & \PolAppleMap & \PolAppleRes & \PolAppleCopy
  & \PolAppleChoice & \PolAppleObserved \\
Ryzen APU & \PolApuMap & \PolApuRes & \PolApuCopy
  & \PolApuChoice & \PolApuObserved \\
GH200, \GhBigGB\,GB & \PolGhMap & n/a & \PolGhCopy
  & \PolGhChoice & \PolGhObserved \\
RTX 5070 Ti & \PolDiscMap & \PolDiscRes & \PolDiscCopy
  & \PolDiscChoice & \PolDiscObserved \\
\bottomrule
\end{tabular}
\caption{Algorithm~\ref{alg:endpoint} applied to the measured
deployments, each at its deployed working set $W$
(\KthreeSpineGB\,GB for the M5~Max spine;
\S\S\ref{sec:apu}--\ref{sec:discrete} for the rest, with
\GhBigGB\,GB on GH200).
\emph{Rule} is the endpoint Algorithm~\ref{alg:endpoint}
returns (ties resolve to map); \emph{outcome} is what the deployment
did: \emph{adopted} where mapping served the weights,
\emph{tied} where mapping and streaming fell within the band,
\emph{rejected} where a one-time copy was retained.  The GH200
\emph{copy} cell is that platform's overlapped-streaming arm
(\GhOverlap\,GB/s), its deployable staging path.}
\label{tab:policy}
\end{table}

\subsection{The held-out machine}\label{sec:heldout}
The endpoint procedure runs twice on the \AirRam\,GB held-out
\AirChip\ machine over a \AirModel\ checkpoint whose \AirMatrices\
matrices stream \AirW\,GB per token, in two memory states that
exercise different branches.  Session one uses the internal
extension's producer.  Its map, resident, and per-use-copy probes
measure \AirMap, \AirResident, and \AirCopy\,GB/s, and resident
storage coexists with the mapping after one pass of every arm in only
\AirResFits\ processes, so the capacity gate removes resident;
mapping is faster than per-use copy.  Session two uses the standalone
producer with more headroom: the probes measure
\AirDmMap/\AirDmResident/\AirDmCopy\,GB/s with coexistence in
\AirDmResFits\ processes,
and the leaders fall inside the \PolTieBand\% band, which resolves to map.
The probe rates select the endpoint before the corresponding
end-to-end run, and mapping is the faster
of the two endpoints measured end to end: \AirPublicTps\ and
\AirDmPublicTps\,tok/s through adoption versus \AirCopyTps\ and
\AirDmCopyTps\ through per-use copy, identical tokens in both
sessions.  The resident endpoint was not run end to end in either
session.

Under the matched 1.06\,GB-pass protocol of \S\ref{sec:tax}, rather
than the endpoint probe over $W$, the machine measures 88.0, 88.6, and
88.9\,GB/s for the mapped, resident, and pure-read arms, with median
paired differences of $+0.3$ and
$+0.5$\,GB/s (below 1\%)---the M5~Max pattern at smaller scale.
Under a single-warm-up protocol, total time per 1.06\,GB pass
preserves the ordering of Fig.~\ref{fig:ordering}: in-stream 38\,ms,
plain event 49, shared event 55, and host drain 124.

The larger \AirBigW\,GB working set fails the residency precondition:
after the framework copy is created, \AirEvictAfter\% of the
mapping stays cached and every path approaches storage throughput.  The
machine does not test a case where resident storage fits reliably and is
materially faster than mapping.

\finding{RQ4}{On the held-out machine, the endpoint procedure selects
mapping in both sessions---one exercises the capacity gate, the other
the tie band---and mapping is the faster of the two endpoints measured
end to end: \AirPublicTps/\AirDmPublicTps\ versus
\AirCopyTps/\AirDmCopyTps\,tok/s for map versus per-use copy,
token-identical, gaps of \AirAdoptRatio--\AirDmAdoptRatio$\times$.
The machine's paired rates and ordering hierarchy also match the
M5~Max pattern.  Neither session ran the resident endpoint end to end.}

\subsection{What frameworks should package}
The kernels and interchange mechanisms already exist; the missing
surface is a mapped-file constructor that handles alignment, lifetime,
mutability, and placement policy: a read-only mapped storage object
that propagates immutability and selects adoption only when the pages
inhabit a GPU-readable domain.  Packed kernels remain a separate concern.  Adoption removes the copy;
a framework that lacks the checkpoint's arithmetic format still performs
dtype expansion.

Once a capsule becomes ordinary framework
storage, downstream operations inherit the framework's stream and dependency
rules, as the PyTorch, MLX, and llama.cpp paths demonstrate.  In that
packaging, the capacity check measures hot-set residency and
representation count rather than model size against installed memory;
no scheduler or intermediate-representation changes are required.

\section{Related work}\label{sec:related}

\textbf{Mapped-weight and native runtimes.}
llama.cpp/ggml includes mmap-backed model loading and a Metal backend that
controls its own storage and command pipeline \citep{llamacpp}.  BaseRT is a
native Metal runtime with custom kernels and dispatch logic \citep{basert},
while a WebAssembly prototype demonstrates no-copy sharing of
externally allocated memory with Metal \citep{wasmmetal}.  These systems
illustrate direct control of storage or execution outside a general tensor
framework's loader.  MLX documents conditional zero-copy DLPack import for
reusable, non-private Metal buffers \citep{mlxdlpackdoc}.  Our producer
supplies the distinct step of constructing a framework-importable tensor from
a file mapping and evaluates the execution conditions under which that route
is useful.  An MLX prototype modified the Metal allocator to back buffers with
file mappings; alignment constraints led to one temporary file per tensor,
and performance degraded once page-cache eviction introduced storage I/O
\citep{mlx615}.  Low-RAM weight streaming remains a live upstream
request in MLX and mlx-lm~\citep{mlx2878,mlxlm1438} and for
unified-memory llama.cpp~\citep{lcpp21827}.

\textbf{Offload, sparsity, and serving.}
LLM in a Flash, PowerInfer, Deja Vu, and MoE offload systems reduce or
schedule the bytes moved \citep{llminaflash,powerinfer,dejavu,moeinfinity,fiddler,
zeroinf,powerinfer2}.  FlexGen and DeepSpeed-Inference schedule transfers
among storage, CPU memory, and discrete-GPU memory
\citep{flexgen,deepspeed}.  ServerlessLLM accelerates checkpoint loading
\citep{serverlessllm}; continuous batching and paged attention improve
resident serving \citep{orca,vllm,sarathi}.  Adoption is complementary: it
changes whether cached bytes acquire a second representation before a kernel
reads them.  A companion study evaluates kernel-managed
expert residency for trillion-parameter MoE serving, one layer above
the read path studied here~\citep{expertcache}.

\textbf{Integrated and coherent memory.}
Zero-copy shared memory on integrated GPUs is longstanding
\citep{intelzc,hashjoin,apu2015,dashti}, and a recent study
characterizes shared-virtual-memory designs across diverse
applications~\citep{svm24}.  CUDA unified-memory work
focuses on migration, prefetch, and oversubscription on discrete devices
\citep{unifiedmemsurvey,uvmoversub,griffin,batchaware,cudaum}.  Grace--Hopper
provides coherent CPU--GPU access across NVLink-C2C, and prior work shows that
performance depends strongly on placement \citep{gh200move}.  SuperInfer uses
the same class of coherent link for KV-cache movement and scheduling
\citep{superinfer}.  Our cross-architecture result separates file-backed
direct-read placement from anonymous migration and evaluates when the same
mechanism should be enabled or rejected.

\textbf{Memory mapping and interchange.}
The value of \texttt{mmap} for fast storage depends on workload and eviction
policy \citep{mmapio,fastmap,pagecachestudy}.  PyTorch exposes memory-mapped
CPU tensors through \texttt{torch.from\_file} \citep{torchfromfile}; moving
such a tensor to an accelerator remains a separate placement operation.
DLPack standardizes producer-owned exchange and includes the Metal device type
used here \citep{dlpack}.  Our producer makes the file mapping itself
accelerator storage, and we evaluate the placement policy and ordering
conditions that decide when to do so.

\section{Scope and limitations}\label{sec:limits}
Platform coverage is asymmetric.  Full framework integrations exist for
Metal on Apple silicon and Vulkan on one AMD APU; GH200 is evaluated
with placement probes and storage rebinding rather than a production
loader; the discrete-GPU experiment is a control.  The APU integration
comprises \ApuRuns\ paired processes; the GH200 evaluation spans
four runs across three instances.  We have not
measured a broader range of integrated and coherent-memory devices.

The workload is low-batch decode.  The ratio measured in the $T$-sweep
of \S\ref{sec:scale} is $1+C(T)/A(T)$, where $A(T)$ is the adopted
arm's arithmetic-and-dispatch time and $C(T)$ the stock arm's
additional ingestion time.  A faster kernel under
the same storage path can only raise the ratio, so the measured ratios are
lower bounds for that substitution; they do not extend to batched GEMM,
prefill, or concurrent serving, where layout and overlap change $C$.

Adoption has two preconditions: page-aligned tensor ranges (an aligned
container or a one-time relayout) and an active mapped set that stays
in the page cache.  Once residency fails, storage traffic dominates and
the in-memory endpoint comparison no longer applies.

\section{Conclusion}

File-backed weights can already sit in memory the GPU can read, yet a
framework loader may copy them into a second allocation before every
use.  A mapped-file DLPack producer removes that ingestion tax under a
three-condition execution contract (adopt the pages, keep activations
on the accelerator, order on the GPU), reaching resident-class
bandwidth on Apple silicon, improving an integrated Vulkan runtime,
and falling inside the selection band against overlapped streaming on
a coherent link.  On a discrete GPU the same path is \LcppDiscSlowdown$\times$
slower and should be rejected.  The deployment rule is topology- and
working-set-aware: adopt cached pages only when they already sit in a
GPU-readable domain.

\clearpage
\bibliographystyle{ACM-Reference-Format}
\bibliography{references}
\end{document}